\documentclass[%
aip,
 amsmath,amssymb,
 reprint,%
]{revtex4-1}

\usepackage{graphicx}
\usepackage{dcolumn}
\usepackage{bm}

\usepackage[utf8]{inputenc}
\usepackage[T1]{fontenc}
\usepackage{mathptmx}
\usepackage{siunitx}
\usepackage{hyperref}

\newcommand*{\symb}[1]{\ensuremath{\mathrm{#1}}}
\newcommand*{\VTCNE}{\ensuremath{\mathrm{V[TCNE]}_x} }
\newcommand*{\VTCNEs}{\ensuremath{\mathrm{V[TCNE]}_x}}
\newcommand*{\AlOxs}{\ensuremath{\mathrm{AlO}_x} }
\newcommand*{\AlOx}{\ensuremath{\mathrm{AlO}_x}}
\let\oldnum\num
\renewcommand*{\num}[1]{\oldnum[separate-uncertainty]{#1}}
\let\oldSI\SI
\renewcommand*{\SI}[1]{\oldSI[separate-uncertainty]{#1}}

\begin{document}


\title{Low-Damping Ferromagnetic Resonance in Electron-Beam Patterned,
  High-\(Q\) Vanadium Tetracyanoethylene Magnon Cavities}

\author{Andrew Franson}
\affiliation{Department of Physics, The Ohio State University, Columbus, Ohio 43210, USA}
\author{Na Zhu}
\affiliation{Department of Electrical Engineering, Yale University, New Haven, Connecticut 06511, USA}
\author{Seth Kurfman}
\author{Michael Chilcote}
\affiliation{Department of Physics, The Ohio State University, Columbus, Ohio 43210, USA}
\author{Denis R. Candido}
\affiliation{Department of Physics and Astronomy, University of Iowa, Iowa City, Iowa 52242, USA}
\affiliation{Pritzker School of Molecular Engineering, University of Chicago, Chicago, Illinois 60637, USA}
\author{Kristen S. Buchanan}
\affiliation{Department of Physics, Colorado State University, Fort Collins, Colorado 80523, USA}
\author{Michael E. Flatt\'e}
\affiliation{Department of Physics and Astronomy, University of Iowa, Iowa City, Iowa 52242, USA}
\affiliation{Pritzker School of Molecular Engineering, University of Chicago, Chicago, Illinois 60637, USA}
\author{Hong X. Tang}
\affiliation{Department of Electrical Engineering, Yale University, New Haven, Connecticut 06511, USA}
\author{Ezekiel Johnston-Halperin}
\affiliation{Department of Physics, The Ohio State University, Columbus, Ohio 43210, USA}
\email{johnston.halperin@gmail.com}

\date{\today}

\begin{abstract}
    Integrating patterned, low-loss  magnetic materials into microwave
    devices and circuits presents many  challenges due to the specific
    conditions that  are required  to grow ferrite  materials, driving
    the   need   for   flip-chip  and   other   indirect   fabrication
    techniques. The  low-loss (\(\alpha = \)  \num{3.98 \pm 0.22e-5}),
    room-temperature  ferrimagnetic   coordination  compound  vanadium
    tetracyanoethylene (\VTCNEs) is a promising new material for these
    applications  that is  potentially  compatible with  semiconductor
    processing.   Here  we  present the  deposition,  patterning,  and
    characterization  of \VTCNE  thin  films  with lateral  dimensions
    ranging  from   1  micron  to  several   millimeters.   We  employ
    electron-beam   lithography   and   liftoff  using   an   aluminum
    encapsulated      poly(methyl      methacrylate),      poly(methyl
    methacrylate-methacrylic acid) copolymer bilayer (PMMA/P(MMA-MAA))
    on sapphire and silicon. This process can be trivially extended to
    other common  semiconductor substrates.  Films patterned  via this
    method maintain  low-loss characteristics down to  25 microns with
    only a factor of 2 increase down to 5 microns. A rich structure of
    thickness  and  radially  confined  spin-wave  modes  reveals  the
    quality of  the patterned films. Further  fitting, simulation, and
    analytic analysis  provides an  exchange stiffness, \(A_{ex}  = \)
    \SI{2.2  \pm 0.5e-10}{erg\per\cm},  as well  as insights  into the
    mode  character and  surface spin  pinning.  Below  a micron,  the
    deposition  is  non-conformal,  which  leads  to  interesting  and
    potentially useful  changes in  morphology. This  work establishes
    the  versatility  of  \VTCNE  for  applications  requiring  highly
    coherent magnetic excitations ranging from microwave communication
    to quantum information.
\end{abstract}


\maketitle 





Interest  in low-loss  magnetic thin  films  has been  growing due  to
potential applications in magnonics and quantum information as well as
the   potential    for   compact,    high-efficiency   magnetoelectric
devices.\cite{Cheng_2018, Nikitin_2015,  Chumak_2017} In the  field of
magnonics and spintronics,  yttrium iron garnet (\symb{Y_3Fe_5O_{12}},
YIG), an  electrically insulating ferrite that  exhibits extremely low
Gilbert damping, \(\alpha \approx\) \num{6e-5}  and a linewidth of 3.4
G  at 9.6  GHz for  pristine nanometer-thick  films, is  currently the
leading            material           for            magnetoelectronic
circuits.\cite{Houchen_Chang_2014,   Hauser_2016}    The   low-damping
present in YIG  films has led to its  incorporation in magnetoelectric
circuits and it also plays a  prominent role in the study of magnonics
research.\cite{Serga_2010,  Schneider_2008, Zahwe_2010,  Nikitin_2015,
  Chumak_2019} Patterning of YIG, however, presents a challenge.  When
patterned, the  damping increases to \(\alpha  \approx\) \num{4e-4} to
\num{8.74e-4} for  ion-milled films,\cite{Hahn_2014, Jungfleisch_2015}
and \(\alpha  \approx\) \num{2.9e-4}  to \num{5e-4}  for liftoff-based
films.\cite{Krysztofik_2017,    Li_2016,     Zhu_2017}    Furthermore,
post-growth   annealing  steps   at  temperatures   as  high   as  850
\si{\celsius}  are generally  required to  attain even  these degraded
damping values,  and the lowest  damping values are only  achieved for
films deposited  on the lattice-matching substrate  gadolinium gallium
garnet  (\symb{Gd_3Ga_5O_{12}}, GGG),  both  of  which provide  strict
limits      on      direct       integration      with      functional
devices.\cite{Onbasli_2014,  Manuilov_2010,   Manuilov_2009}  Vanadium
tetracyanoethylene (\VTCNEs,  \symb{x \approx 2}), on  the other hand,
is  a low-loss  (sub-Gauss  linewidth at  9.83 GHz),  room-temperature
(\symb{T_c=}~600~K)  ferrimagnet that  can be  deposited optimally  at
\SI{50}{\celsius}  and \SI{30}{\mmHg}  without  the  need for  lattice
matching.\cite{Harberts_2015,   Yu_2014}   These   relatively   benign
deposition  conditions  allow for  deposition  on  a wide  variety  of
substrates  and  pre-patterned  circuits,  positioning  \VTCNE  as  an
exciting   option   for   on-chip   magnetic   and   magnonic   device
incorporation.\cite{Zhu_2016,   Liu_2018,  chilcote19:spin}   However,
realizing this promise has been limited  by the lack of techniques for
patterning \VTCNE films at micron to sub-micron length scales.

Here we  present a method  for depositing and patterning  \VTCNE using
standard electron-beam lithography techniques with additional steps to
preserve its high \symb{T_c} and low-loss characteristics. The primary
hurdles to  micron-scale pattering  of \VTCNE  are its  sensitivity to
oxygen and solvents traditionally used  in fabrication.  Our past work
has  addressed  air-sensitivity  via  encapsulation  in  a  commercial
organic light-emitting diode (OLED)  epoxy, increasing its lifetime in
air  from  hours to  months,\cite{Froning_2015}  and  there are  other
commercial options,  such as potting,\cite{Bardoliwalla}  that promise
to protect films indefinitely.  This leaves solvent sensitivity, which
inhibits the use of traditional patterning techniques for two reasons:
i) the presence of solvent in the resist layer inhibits the deposition
of  \VTCNE as  the solvent  outgasses during  growth, and  ii) liftoff
requires a  solvent soak that will  in general destroy or  degrade the
CVD-grown \VTCNE  film. Here  we address both  of these  challenges by
using a thin \AlOxs encapsulating layer for the resist and identifying
a \VTCNEs-compatible solvent, respectively, demonstrating micron-scale
patterning  of \VTCNE  films with  no apparent  increase in  microwave
loss. The patterned structures  are characterized by scanning electron
microscopy (SEM) and by a combination of ferromagnetic resonance (FMR)
and   comparison   with   micromagnetic   simulations   and   analytic
calculations.

The  CVD  thin-film  growth  process typically  results  in  a  smooth
blue-black  \VTCNE film  uniformly  distributed  across the  substrate
surface.  When a  resist is  applied to  the substrate  before growth,
however, \VTCNE  deposition results  in non-uniform coverage  and poor
\VTCNE quality. This  is attributed to chemical  reactions between the
released solvents and the  precursors (tetracyanoethylene and vanadium
hexacarbonyl).  Solvents  present  in common  resists  including  LOR,
MICROPOSIT  S1800  series,  and poly(methyl  methacrylate)  result  in
macroscopically inconsistent deposition of  \VTCNE across the resist’s
surface as  well as inside patterned  areas. In order to  address this
solvent  sensitivity, a  \SI{3}{\nm}  layer of  aluminum is  thermally
deposited  after  development to  encapsulate  the  resist layer.  The
aluminum is then oxidized with a ten-minute ultraviolet ozone clean in
a UVOCS T10x10/OES prior to \VTCNE deposition.

In  prior  \VTCNE precipitation  synthesis  studies,\cite{Thorum_2006,
    Pokhodnya_2001, Zhang_1996}  several solvents  have been  shown to
precipitate  \VTCNE with  a modest  impact  on the  \symb{T_c} of  the
resulting powder. Since  dichloromethane has a small  impact on \VTCNE
quality  and readily  dissolves poly(methyl  methacrylate) (PMMA)  and
poly(methyl methacrylate-methacrylic  acid 8.5\%) (P(MMA 8.5  MAA)) at
room temperature,  this is the  resist-solvent pair chosen  to address
the  challenge  of  solvent  based liftoff.  Specifically,  this  work
focuses on  495PMMA A6  on MMA  (8.5) MAA  EL 11  as a  resist bilayer
(PMMA/P(MMA  8.5  MAA)) to  additively  pattern  low-loss \VTCNE  onto
sapphire, with  the understanding that this  patterning process should
trivially  extend  to   other  inorganic  substrates.\cite{Tseng_2003,
    Tennant_2016}


\begin{figure}
    \centering
    \includegraphics{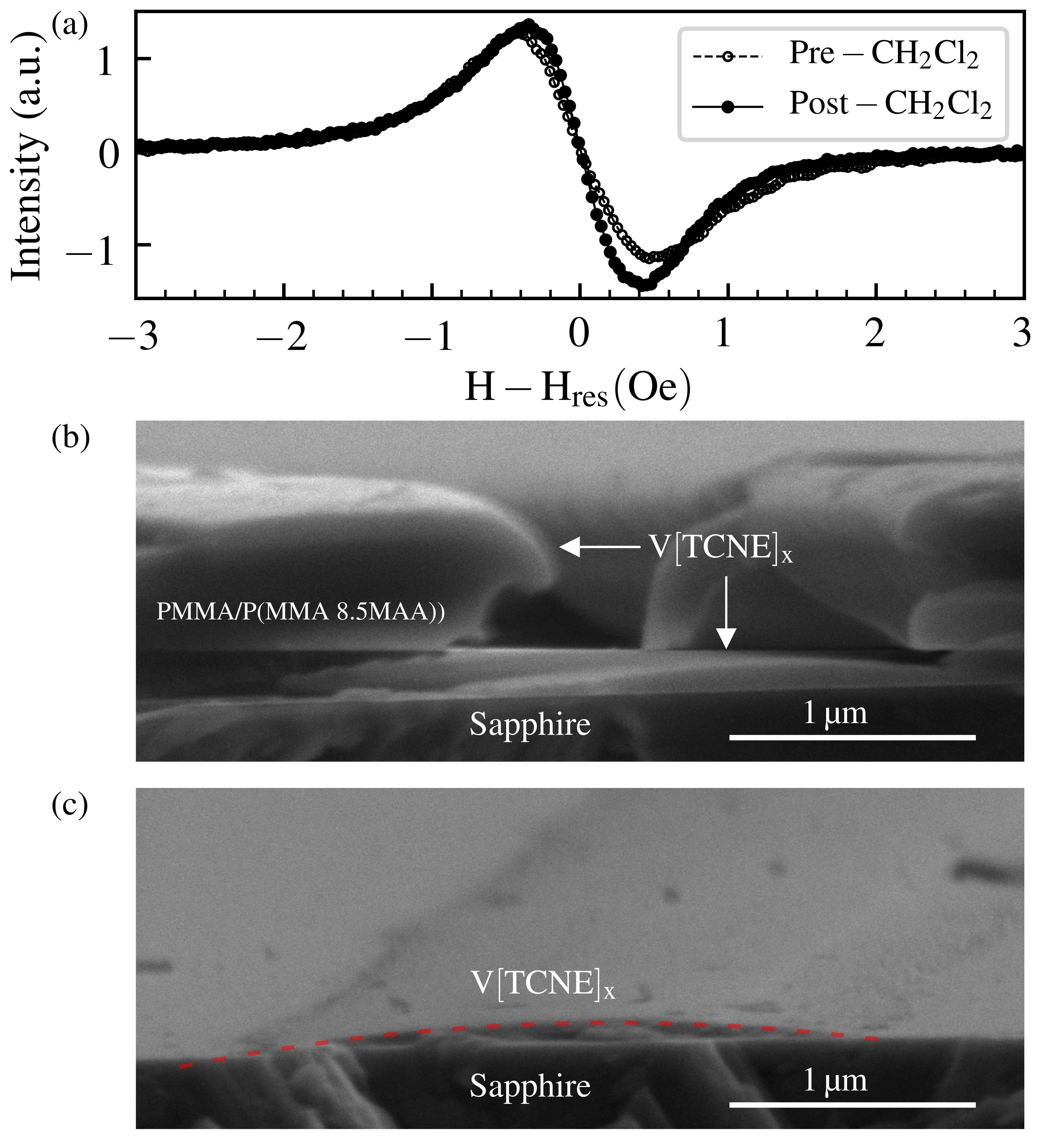}
    \caption{(a) \VTCNE thin  film FMR at 9.83 GHz  and resonant field
        of 3565 Oe before and after a 2.5-hour soak in dichloromethane
        (\symb{CH_2Cl_2}).  SEM   images  of  patterned   \VTCNE  film
        morphology   before   (b)   and   after   (c)   liftoff   with
        dichloromethane reveal a  parabolic deposition morphology into
        a  \SI{2}{\um} wide  channel. The  parabolic cross-section  is
        highlighted with a dashed red line in (c).}
    \label{fig:ch2cl2}
\end{figure}

Figure~\ref{fig:ch2cl2}(a) shows  the FMR  response at  9.83 GHz  of a
\VTCNE thin film  before and after a 2.5-hour  soak in dichloromethane
in       a       nitrogen       atmosphere       (<~10~ppm~\symb{0_2},
<~2~ppm~\symb{H_2O}). The linewidth and lineshape of the resonance are
largely unchanged, indicating that there is little to no incorporation
of dichloromethane  into the  CVD-grown film  on that  timescale.  The
linewidth narrows slightly, possibly due to changes in the ordering of
the \VTCNE  due to solvent annealing.\cite{Hu_2014}  The \VTCNE growth
morphology that results from the above process is characterized by SEM
and  is shown  in Fig.~\ref{fig:ch2cl2}(b-c).   Unlike physical  vapor
deposition,  CVD deposition  is driven  by a  combination of  flow and
diffusion. Figure~\ref{fig:ch2cl2}(b) shows  how the \VTCNE deposition
is limited by the flow characteristics through the patterned features.
In  particular, \VTCNE  does not  form vertical  sidewalls but  rather
forms gently  sloped sidewalls  at an  angle of  about \SI{6}{\degree}
over a distance  of approximately a micron from the  edge.  This leads
to a parabolic profile, as one  would expect from the velocity profile
resulting     from     laminar     flow     through     a     channel,
Fig.~\ref{fig:ch2cl2}(c).  These results suggest that there are likely
opportunities  to tune  the structure  profile by  controlling channel
width,  flow direction,  resist height,  and resist  morphology.  This
cross-sectional profile  is difficult  to achieve with  other material
systems and  deposition techniques,\cite{Tait_1993}  and it  may prove
beneficial for studies  of spin-wave confinement as it  offers a means
to realize an approximation of an adiabatic boundary.

\begin{figure}
    \centering
    \includegraphics{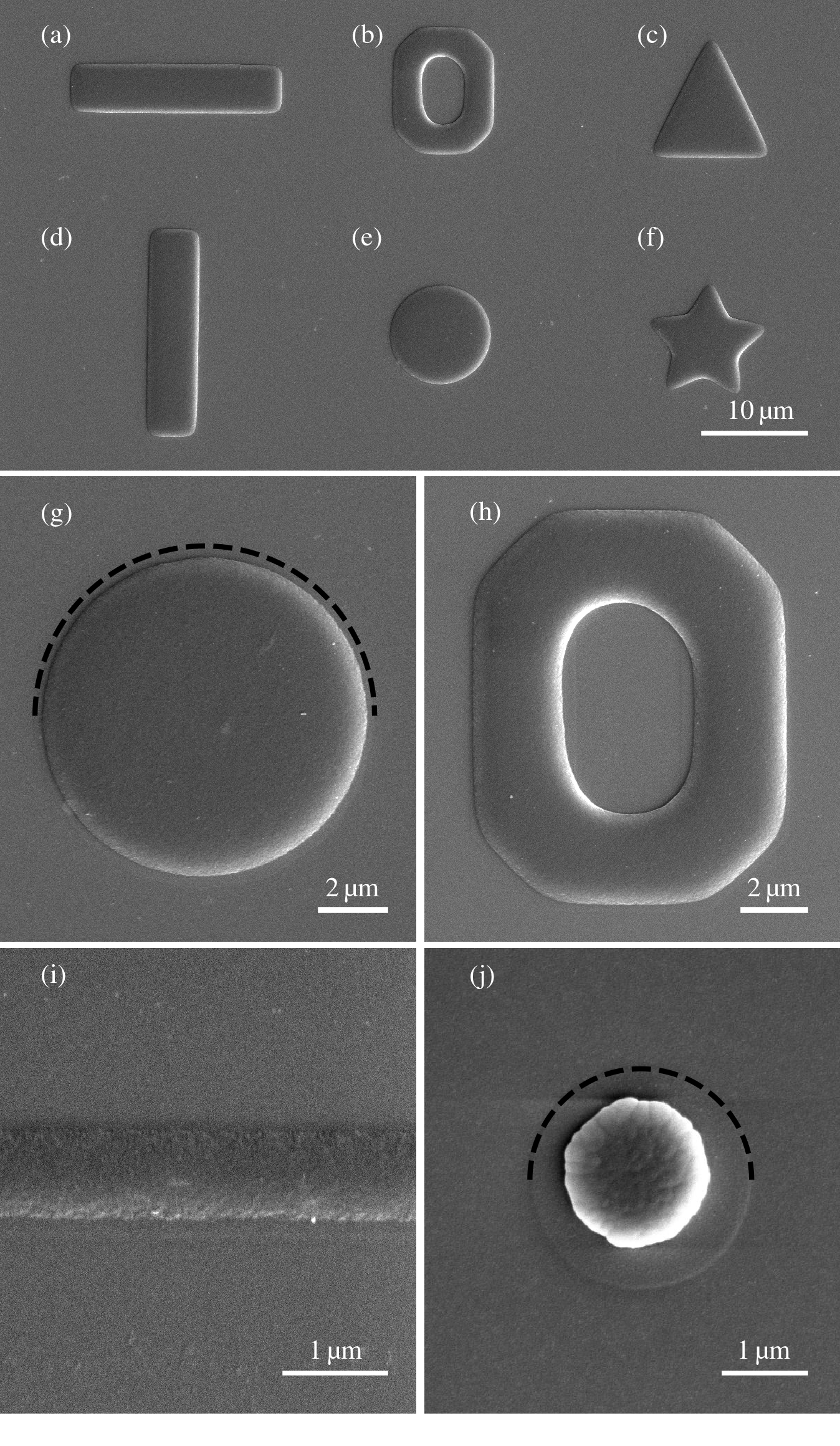}
    \caption{Top view  SEM images of  various \VTCNE patterns  after a
      1.0 hr growth, about 300  nm thick.  (a-f) Deposition morphology
      of several shapes with various  features ranging from concave to
      convex  to  antidot.   (g-h)  Enhanced views  of  (e)  and  (b),
      respectively, show  that flow-induced  anisotropy is  present in
      the complex  Block O shape  (h). (i-j)  For features of  order 1
      micron  or smaller,  the restricted  flow begins  to affect  the
      pattern deposition.   The black, dashed semi-circles  in (g) and
      (j)  highlight  the  \AlOxs  profile that  is  present  for  all
      shapes.}
    \label{fig:topview}
\end{figure}

Flow-limited deposition can, in principle,  also lead to anisotropy in
patterned features based  on alignment between the  flow direction and
internal structure.  Figure~\ref{fig:topview}  shows various patterned
structures of \VTCNE  that are designed to explore  these effects. The
images  show that  there  is  anisotropic growth  for  several of  the
structures. The deposition  time for these samples is  1.0 hr, leading
to   a    nominal   thickness    of   300    nm.    All    shapes   in
Fig.~\ref{fig:topview}  have   a  faint   outline  that   reveals  the
ballistically        deposited         \AlOxs        layer.         In
Figs.~\ref{fig:topview}(g,~j),  a dashed  black  semi-circle has  been
superimposed over the \AlOxs outline to  make it easier to compare the
\AlOxs and \VTCNE  morphology. The outline shows that  there is little
to  no offset  or ellipticity  present in  the patterned  \SI{10}{\um}
diameter  \VTCNE  disk.  Figure~\ref{fig:topview}(h),  however,  shows
significant  anisotropy as  measured  by the  differences between  the
faint \AlOxs outline and the \VTCNE pattern. The flow direction across
the shape  is left to right  with a \SI{20}{\degree} tilt  towards the
top.  The flow direction manifests in a more complete, laminar profile
along the top and bottom whereas eddies inhibit deposition in the left
and right  interior edges. Laminar flow  over a step predicts  an eddy
approximately  as   wide  as  the  step   is  tall.\cite{Shankar_2000,
  Taneda_1979} This is consistent with the  fact that the 540 nm thick
PMMA/P(MMA 8.5 MAA))  bilayer results in a  roughly \SI{500}{\nm} wide
region of reduced flow which leads to a taper in the morphology.  This
effect is also seen in Figs.~\ref{fig:topview}(c, f) where the concave
features  at  the  corners  see  a  reduction  in  deposition  roughly
\SI{800}{\nm} away from the planned shape shown by each \AlOxs peak.

Figures~\ref{fig:topview}(i, j) further explore  the impacts of length
scale on  gas flow  and growth  morphology using  bars and  disks. The
\SI{1}{\um} wide  bar in  Fig.~\ref{fig:topview}(i) acts as  a channel
for  gas flow,  yielding  a parabolic  deposition  profile similar  to
Fig.~\ref{fig:ch2cl2}(b, c) across  the shape and good  filling of the
ballistic    profile   when    the   thickness    of   the    bar   in
Fig.~\ref{fig:topview}(i)  is about  200 nm  thick.  In  contrast, the
\SI{1.77}{\um} diameter  disk in Fig.~\ref{fig:topview}(j)  is visibly
off-center, with  a 100 nm offset  towards the top-left of  the \AlOx.
The deposition  morphology in Fig.~\ref{fig:topview}(j)  resembles the
eddy flow  structure from  boundary-driven flow into  a cavity  with a
depth-to-width  ratio  of one-third,  suggesting  that  flow over  the
features   is  laminar   and  the   resulting  deposition   shape  and
cross-section   can   be   simulated   from   flow.\cite{Shankar_2000,
  Taneda_1979} It may be possible to achieve smaller features by using
thinner  resist  layers,  or   by  choosing  pattern  geometries  that
intentionally channel the  flow, but these approaches  will be pattern
specific and are beyond the scope of this work.

To explore the  utility of this patterning technique  for magnonic and
magnetoelectric   devices,  the   magnetization   dynamics  of   these
microstructures are  studied using FMR. Measurements  are performed at
room  temperature in  a  Bruker EPR  spectrometer  with the  microwave
frequency held near 9.83 GHz while the applied magnetic field is swept
across  the \VTCNE  resonance. Scans  are then  repeated for  multiple
polar   angles,  \(\theta\),   from   out-of-plane   (\(\theta  =   \)
\SI{0}{\degree}, OOP) to in-plane (\(\theta = \) \SI{90}{\degree}, IP)
and for multiple  azimuthal angles, \(\varphi\), from  parallel to the
\(x\)\nobreakdash-axis   (\(\varphi    =   \)    \SI{0}{\degree})   to
perpendicular          to          the          \(x\)\nobreakdash-axis
(\(\varphi =\)~\SI{90}{\degree}).

\begin{figure}
  \centering
  \includegraphics{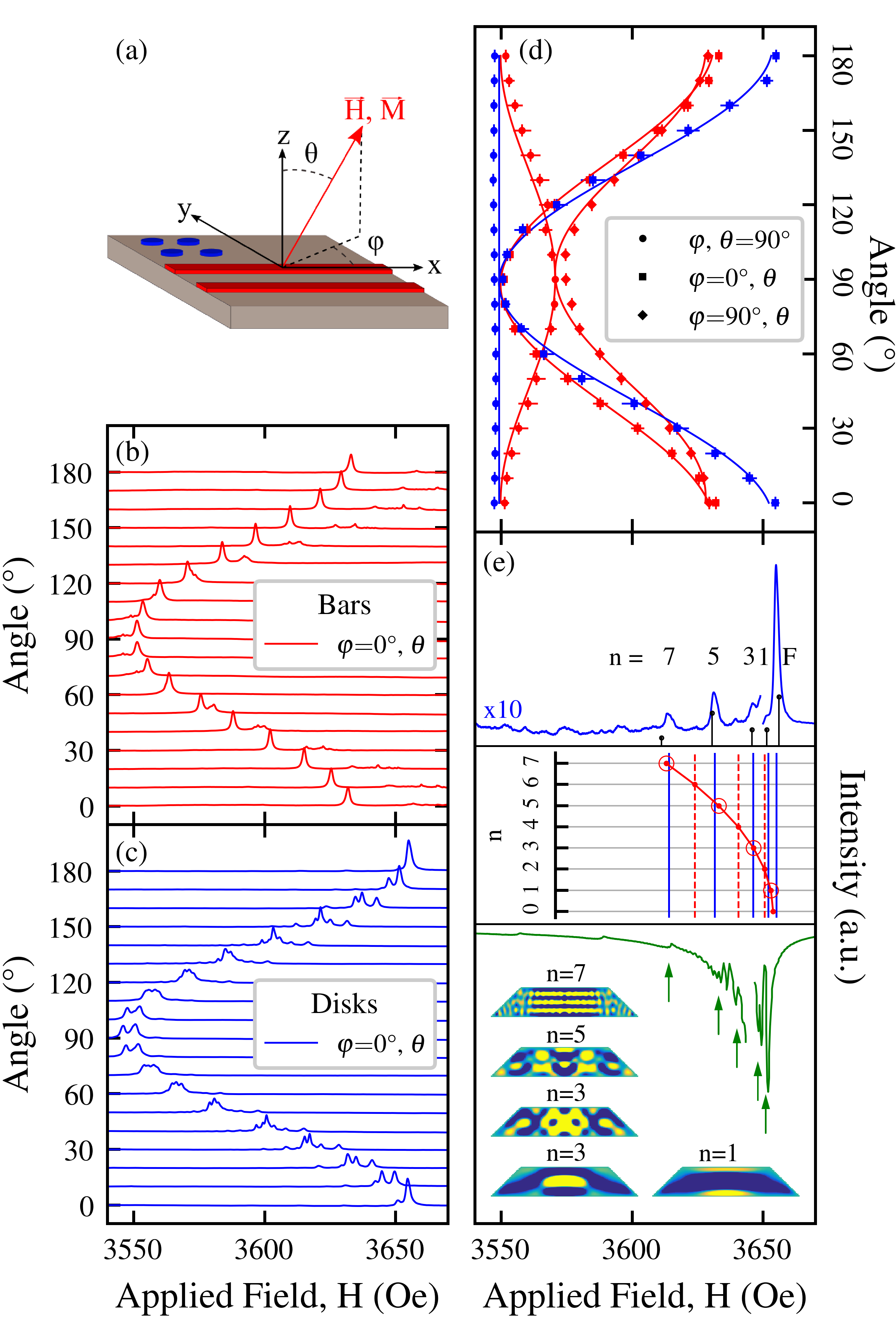}
  \caption{Ferromagnetic   resonance  characterization   of  patterned
    \VTCNE bars and  disks. (a) Schematic of  the measurement geometry
    for  the array  of \SI{1}{\um}  bars (red)  and \SI{5}{\um}  disks
    (blue)  of \VTCNEs.   (b-c) FMR  response vs.   applied field  for
    \(\theta=0\)\si{\degree}   to    \(\theta=180\)\si{\degree}   with
    \(\varphi\)  fixed along  the \(x\)-axis.   (d) Uniform  mode peak
    positions  (symbols)   for  three  high-symmetry   directions  and
    corresponding   fit   curves,   Eq.   3.    (e)   Comparisons   of
    experimental, analytic,  and simulated OOP FMR  spectra.  OOP disk
    linescan (top,  blue) with analytically calculated  peak positions
    (top,   black),   thickness-mode   fitting  (middle,   red),   and
    micromagnetic  simulations (bottom,  green) with  cross-sectional
    mode maps  (bottom left, yellow  and dark blue  represent positive
    and negative motion, respectively, at  an instant of time).  Solid
    blue  lines  in  middle  panel  indicate  experimentally  observed
    odd-mode resonance fields and the  symbols show the best fit field
    values  where the  circled points  (odd modes)  were used  for the
    fits.  Red dashed lines show predicted even-mode resonances.}
  \label{fig:fmrcenters}
\end{figure}

Figure~\ref{fig:fmrcenters} shows the  results of FMR characterization
of    \SI{1}{\um}    wide    bars     aligned    parallel    to    the
\(x\)\nobreakdash-axis and  \SI{5}{\um} diameter  disks. The  bars are
spaced \SI{20}{\um} center  to center in a 1D array  and the disks are
spaced  \SI{40}{\um} center  to center  in  a 2D,  square array.   The
position of  the uniform mode  of the bars  (red) and disks  (blue) is
tracked as a function  of orientation in Figs.~\ref{fig:fmrcenters}(b,
c).  The bars show a single-peaked  resonance that varies from 3550 to
3630 Oe as  the structures are rotated  IP to OOP. The  disks reveal a
more  complicated peak  structure,  that  suggests standing  spin-wave
modes are present, Fig.~\ref{fig:fmrcenters}(i),  and exhibit a larger
field difference  between IP and  OOP resonances.  This  difference is
evident  in  Fig.~\ref{fig:fmrcenters}(d)  where  the  resonances  are
tracked through multiple high-symmetry  directions, revealing the full
anisotropy of these structures.

The  formalism developed  by Smit,  Beljers, and  Suhl\cite{Suhl_1955,
  Smit_1955,  Baselgia_1988}  is used  to  model  this anisotropy.   A
Cartesian  coordinate system  is defined  with \(x\)  parallel to  the
length,  \(y\)  parallel to  the  width,  and  \(z\) parallel  to  the
thickness of  the bars  as shown in  Fig.~\ref{fig:fmrcenters}(a).  By
explicitly considering  the Zeeman and magnetostatic  contributions to
the free  energy, \(F\),  one derives  the expression,\cite{Smit_1954,
  Morrish_2001}
\begin{equation}
  \label{eq:FreeEnergy}
  F = - M_i H_i + \frac{1}{2} M_i N_{ij} M_j,
\end{equation}
where  \(H_i\)  are the  components  of  the applied  magnetic  field,
\(M_i\) are  the components of  the magnetization, and  \(N_{ij}\) are
components  of   the  demagnetizing  tensor,  which   leads  to  shape
anisotropy, with  \(i, j,  k\) defined with  respect to  pattern axes.
Solving for  harmonic solutions  with respect  to time  and minimizing
\(F\) with respect to \(\theta\) and \(\varphi\) yields
\begin{multline}
    \label{eq:omegaExpression}
    \frac{\omega}{\gamma} = 
    \left\{
        [H-4 \pi M_s N_{OP}\cos(2 \theta)] \right. \\
        \times [H-4 \pi M_s N_{OP}\cos^2(\theta)- 4 \pi M_s N_{IP}\cos(2 \varphi)] \\
        \left. -16 \pi^2 M_s^2 N_{IP}^2\cos^2(\theta)\cos^2(\varphi)\sin^2(\varphi) \right\}^{1/2},
\end{multline}
where
\begin{subequations}
  \begin{equation}
    \label{eq:NOP}
    N_{OP} \equiv N_z - N_x \cos^2(\varphi) - N_y \sin^2(\varphi),
  \end{equation}
  \begin{equation}
    \label{eq:NIP}
    N_{IP} \equiv N_x - N_y,
  \end{equation}
\end{subequations}
\(N_x, N_y,  N_z\) are  the diagonal  components of  the demagnetizing
tensor,  \(\theta\)  is  the  polar  angle,  and  \(\varphi\)  is  the
azimuthal angle that  the sample magnetization makes  with the pattern
axes, Fig.~\ref{fig:fmrcenters}(a).  Equation~\ref{eq:omegaExpression}
is  derived  assuming  the  demagnetizing field  is  parallel  to  the
magnetization, so \(N_{ij}  = N_{ij}\delta_{ij} = N_i\),  and that the
magnetization, \((\theta_M,  \varphi_M)\), is parallel to  the applied
field, \((\theta_H, \varphi_H)\), so only  one set of angles is needed
to     describe    the     magnetization     and    applied     field,
\((\theta, \varphi)\).   This approximation  is validated by  the fact
that \(4  \pi M_s\)  for \VTCNE  is less  than 95  G, and  the applied
magnetic fields used for these measurements  are 3500 to 3750 Oe. As a
result, \(\vec{M}\) is  parallel to \(\vec{H}\) to  within 1.5 degrees
for the experiments shown here.

The other  potential source of  anisotropy is the crystal  field which
arises from  the local coordination  of the exchange  interaction. For
uniaxial  crystal-field   anisotropy,  this   crystal  field   can  be
decomposed into components acting along the pattern axes with the same
angular dependence as  the demagnetizing anisotropy. As  a result, the
\(N_i\)     that     are     extracted     from     the     fit     to
Eq.~\ref{eq:omegaExpression} are a  combination of demagnetizing-field
and crystal-field components with the form
\begin{equation}
  \label{eq:EffectiveFields}
  A_{i} \equiv N_{i,extracted} = N_{i,demag} + \frac{H_{i,crystal}}{4 \pi M_s},
\end{equation}
where   \(A_i\)  is   the   observed   anisotropy  tensor   component,
\(N_{i,demag}\) is  the geometric demagnetizing tensor  component, and
\(H_{i,crystal}\) is the additional crystal field along that axis.

Three anisotropy  tensors are  used to determine  the strength  of the
crystal field in the bars and  the disks. The first, \(A_{i,fit}\), is
generated from simultaneous fits to the three red anisotropy curves in
Fig.~\ref{fig:fmrcenters}(d)     from     the     bar     array     to
Eq.~\ref{eq:omegaExpression}.   These  fits yield  \(4  \pi  M_s =  \)
\num{76.57 \pm  1.67} G and  \(|\frac{\gamma}{2 \pi}| =  \) \num{2.742
  \pm 0.04} MHz/Oe, which agree with literature values,\cite{Zhu_2016,
  Yu_2014}    and   \(A_{x,fit}    =   \)    \num{0.0   \pm    0.001},
\(A_{y,fit} = \)\num{0.189 \pm 0.019}, and \(A_{z,fit} = \) \num{0.707
  \pm 0.026}.  The  trace of this anisotropy tensor  is \num{0.896 \pm
  0.046}, indicating  the magnitude of the  crystal-field contribution
is \num{7.96  \pm 2.47} Oe.   Using SEM measurements  to geometrically
determine    a   pure    demagnetizing    tensor    for   the    bars,
\(N_{i,demag}^{bar}\),  that does  not  include crystal-field  effects
yields \(N_{x,demag}^{bar}  = 0\),  \(N_{y,demag}^{bar} =  0.21\), and
\(N_{z,demag}^{bar}    =   0.79\).\cite{Smith_2010}    Comparing   the
\(A_{i,fit}\)  with these  \(N_{i,demag}^{bar}\), the  \(z\) direction
shows the largest difference of 0.09, indicating this crystal field is
oriented along the \(z\)\nobreakdash-axis  of the bars.  The magnitude
of  this crystal  field is  consistent with  previous measurements  of
\VTCNE templated nanowires.\cite{chilcote19:spin} In addition to being
self-consistent, these results also  predict the anisotropy curves for
the disks (blue lines in Fig.~\ref{fig:fmrcenters}(d)).  To test these
fitting results, a final demagnetizing tensor, \(N_{i,demag}^{disk}\),
for         the         disk         are         calculated         as
\(N_{x,demag}^{disk}    =    N_{y,demag}^{disk}    =    0.028\)    and
\(N_{z,demag}^{disk}  =   0.944\)  based   on  SEM   measurements  and
demagnetizing   expressions   from  the   literature.\cite{Kraus_1973}
Combining this with  the \(4 \pi M_s\)  and \(|\frac{\gamma}{2 \pi}|\)
values from the previous fit results in the solid blue curves shown in
Fig.~\ref{fig:fmrcenters}(d) with a combined reduced chi-squared value
of  0.96. Adding  crystal-field effects  degrades the  quality of  the
reduced  chi-squared  value  for \(H_{z,crystal}^{disk}  >  0.7\)  Oe,
indicating the absence  of crystal-field effects in  the disks.  These
results suggest the crystal-field contribution arises from anisotropic
relaxation in the  patterned bars, which corroborates  prior work with
\VTCNE  nanowires  where  an  additional  in-plane  crystal  field  is
reported  due  to  anisotropy  in  the  relaxation  of  the  templated
structures.\cite{chilcote19:spin}

The more complicated spectra of the  disks suggests that the disks are
acting  as spin-wave  cavities with  complex internal  mode structure,
Fig.~\ref{fig:fmrcenters}(e).   Numerical  simulations and  analytical
calculations are carried out to better understand this mode structure.
To begin characterizing the mode structure, the strongest experimental
peaks are compared with the odd analytic thickness modes predicted for
a  thin film  in the  OOP geometry.\cite{Kalinikos_1986}  The vertical
blue lines in  Fig.~\ref{fig:fmrcenters}(e) represent the experimental
peak values.  Fitting to these  peak values using the mode assignments
indicated in Fig.~\ref{fig:fmrcenters}(e)  and the parameters obtained
from the FMR measurements yields the red analytic curve and a value of
the    exchange    stiffness,    \(A_{ex}    =    \)    \SI{2.2    \pm
  0.5e-10}{erg\per\cm}.  The even thickness modes, shown as dashed red
lines, agree  well with  smaller peaks  within the  experimental data.
Analytic     disk      calculations     shown     in      black     in
Fig.~\ref{fig:fmrcenters}(e)  further  describe  the identity  of  the
quantum confined modes and agree well when using this \(A_{ex}\).  The
exchange stiffness depends  on \(M_s\); an approximate  form, found by
several   means,\cite{Skomski_2003,   Skomski_2004,  Moreno_2016}   is
\(A_{ex}\propto    M_{s}^2\).     The   exchange    length    constant
\(\lambda_{ex}  =  \frac{2A_{ex}}{\mu_{0}M_{s}^{2}}\) is  therefore  a
better  metric to  use to  compare samples  with different  saturation
magnetizations.  The difference between  the exchange length from this
study of \(\lambda_{ex} = 9.7 \)  nm and the previously reported value
of 21 nm\cite{Zhu_2016} could be due to differences in grain structure
between the patterned and  unpatterned films\cite{Moreno_2016} as well
as difficulty  in mode  assignment, \(n\), in  prior work  where fewer
modes are visible.

Numeric   modeling  is   performed  using   time-domain  micromagnetic
simulations with the open-source GPU-based software MuMax3 while using
the material parameters  determined from the fits  to the experimental
data.\cite{Vansteenkiste_2014} The factors that have the most relevant
influence on the simulated peak structure are (i) the sloped sidewalls
that (a) have a strong effect on the shape of the lowest frequency set
of peaks which  are comprised of a set of  closely-spaced radially and
lowest-order thickness quantized modes and  (b) apply an overall shift
to the  thickness confined modes,  (ii) the pinning conditions  of the
surfaces  that  have  a  strong   effect  on  the  amplitudes  of  the
thickness-confined   modes,   and   (iii)  the   exchange   stiffness,
\(A_{ex}\),  that controls  the  spacing  between thickness  quantized
modes. Sloped sidewalls  are used in the simulations  to replicate the
shape that occurs due to the slower growth rate within 1 micron of the
resist. The simulations  show that the position of  the most prominent
peak  relative to  the thickness-confined  modes is  sensitive to  the
exact shape of  the sidewalls and the pinning  conditions.  To account
for small differences in the slope of simulated and experimental data,
the higher-order  thickness modes are aligned  with experiment instead
of the uniform mode  in Fig.~\ref{fig:fmrcenters}(e). Simulations with
perfect pinning at  the top and bottom surfaces agree  better with the
experimentally observed  thickness and radial confined  mode structure
as compared to  simulations with top, bottom, or  no pinning; however,
the close agreement between  the calculated even-mode resonance fields
and several smaller  peaks in the experimental  spectrum suggests that
one of the surfaces likely has slightly weaker pinning than the other.
Additional simulations can be found  in the supplement.  The resulting
simulated  frequency  response  of  the  simulation  is  in  green  in
Fig.~\ref{fig:fmrcenters}(e)  along with  several mode  maps at  peaks
indicated by the  green arrows. These maps reveal  quantization in the
thickness  and  radial  directions  in  the  tapered  structure.   The
lower-order thickness  modes each  show distinct  radial quantization.
The \(n=7\) thickness mode, shows a nearly pure thickness quantization
and represents the  sum of multiple closely-spaced  radials modes that
are  excited  simultaneously.   The agreement  between  simulated  and
experimental  spectra demonstrates  control  over  the spin-wave  mode
structure and  lays the  foundation for the  study and  application of
magnon  cavities   with  adiabatic  boundaries  and   engineered  mode
structures.

\begin{figure}
  \centering
  \includegraphics{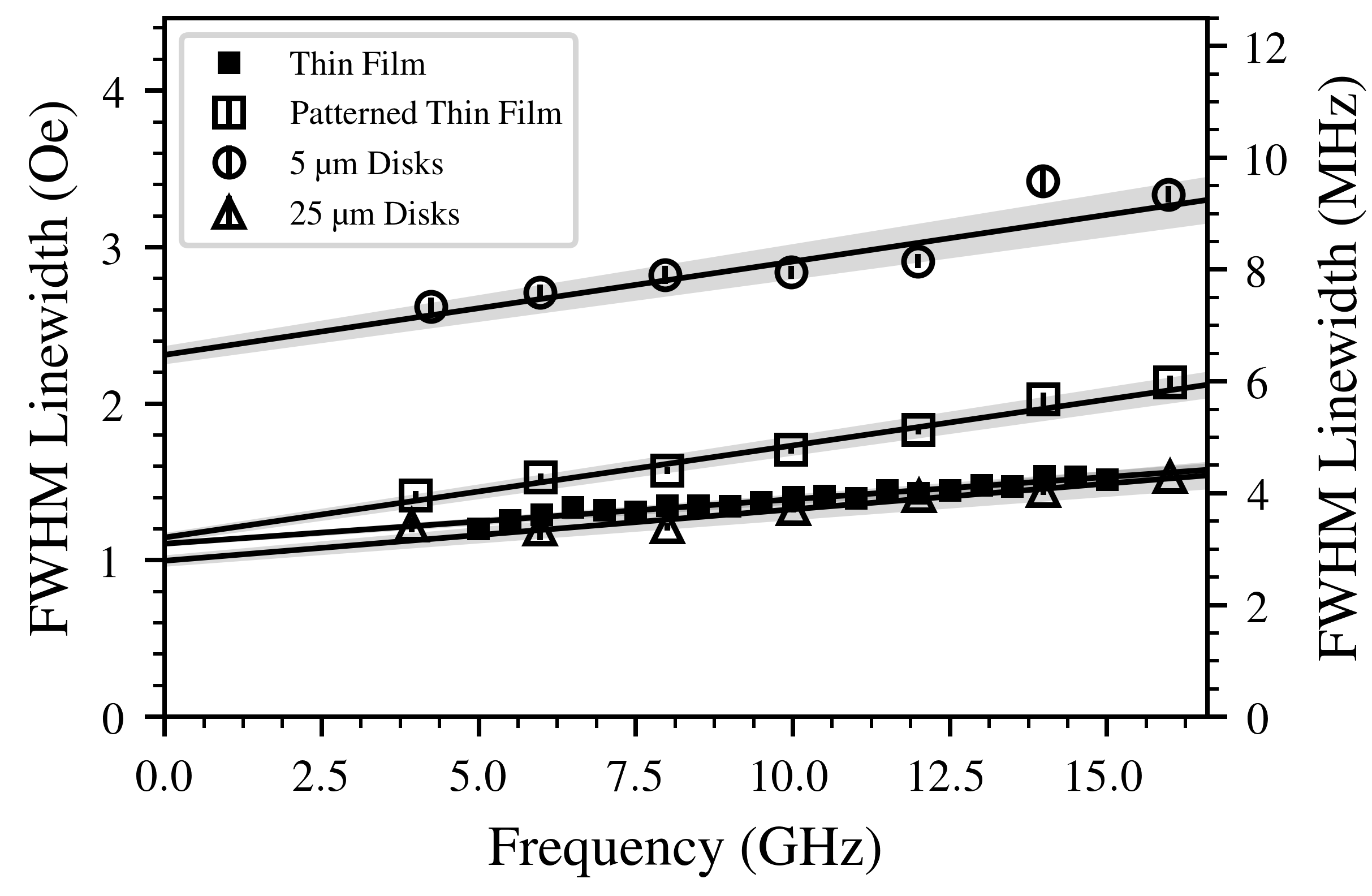}
  \caption{Full-width at half-maximum  linewidth vs resonant frequency
      for various \VTCNE pattern sizes  from thin films to \SI{5}{\um}
      diameter  disks.   All linewidths  are  extracted  from the  OOP
      geometry. All  growths were 1-hour  long, resulting in a  300 nm
      thick film for the \SI{5}{\um} film and 400 nm thickness for the
      rest. The patterned thin film is a  2 mm by 2 mm patterned patch
      of \VTCNEs.}
  \label{fig:gilbert}
\end{figure}

In addition  to analyzing  anisotropy and mode  structure, FMR  can be
used to determine the total magnetic  loss, or damping of these magnon
modes.   This  damping  potentially   contains  both  homogeneous  and
inhomogeneous sources as parameterized via the Gilbert damping factor,
\(\alpha\).\cite{Kalarickal_2006} The damping for the patterned \VTCNE
films  is  measured  via   broadband  ferromagnetic  resonance  (BFMR)
performed  in  a  custom  built microstrip-based  system  wherein  the
applied magnetic  field is held constant  in the OOP geometry  and the
microwave   frequency   is   swept  across   the   \VTCNE   resonance.
Figure~\ref{fig:gilbert}  shows the  linewidth vs  frequency extracted
for  representative  samples  of  disks and  unpatterned  films,  with
vertical lines  indicated the error  in the fits.   Representative raw
data  and fits  can be  found  in the  supplemental information.   The
Gilbert damping is  fit using Suhl's expression for  the full-width at
half-maximum (FWHM) FMR linewidth,\cite{Suhl_1955}
~
\begin{equation}
    \label{eq:SuhlLinewidth}
    \Delta H = \frac{\alpha}{|d \omega_{res} / dH|}
    \frac{\gamma}{M}
    \left( F_{\theta \theta} + \frac{1}{\sin^2 (\theta)} F_{\varphi \varphi}
      \right),
\end{equation}
in      combination      with      phenomenological      inhomogeneous
broadening.\cite{Heinrich_1985} This results in
\begin{equation}
  \label{eq:gilbert}
  \Delta H = \frac{4 \pi \alpha}{|\gamma|} f + \Delta H_0,
\end{equation}
when one uses \(\theta=0\) for the OOP geometry.\cite{Kalarickal_2006}
In Eq.   \ref{eq:gilbert}, \(\Delta H\)  is the FWHM linewidth  of the
resonance, \(\alpha\)  is the Gilbert  damping, and \(\Delta  H_0\) is
the FWHM  contribution from  inhomogeneous broadening. The  fits yield
\(\alpha\) = \num{3.98 \pm  0.22e-5} for unpatterned films, \(\alpha\)
= \num{4.60 \pm  0.44e-5} for \SI{25}{\um} features,  and \(\alpha\) =
\num{8.34 \pm  0.77e-5} for  \SI{5}{\um} disks. The  thin-film damping
result  of  \num{3.98 \pm  0.22e-5}  places  \VTCNE films  comfortably
alongside YIG films as the  lowest magnetic damping material currently
available,  and   the  retention  of  that   ultra-low  damping  after
patterning  is  considerably  better  than  the  reported  values  for
patterned     YIG    structures.\cite{Hahn_2014,     Jungfleisch_2015,
  Krysztofik_2017, Li_2016, Zhu_2017} In  addition to low-damping, the
high-frequency measurements  of the  thin film and  \SI{25}{\um} disks
have Quality  (\(Q\)) factors, \(\frac{f}{\Delta f}\),  of over 3,700,
competitive      with     \(Q\)      factors     for      YIG     thin
films.\cite{Balinskiy_2017} Retaining ultra-low damping and high \(Q\)
in patterned \VTCNE for features as small as \SI{25}{\um} and as large
as millimeters, both  relevant length scales for  many magnonic cavity
applications,\cite{Chumak_2014,     Chumak_2017,     Cornelissen_2018,
  Morris_2017, Zhang_2014, Zhu_2017} combined  with the flexibility to
deposit on  most inorganic substrates, positions  \VTCNE to complement
YIG in magnonic  and magnetoelectric devices where  integration of GGG
or high-temperature  annealing steps  is limiting,  such as  for small
form factors and on-chip integration.

In  summary,  this  work  demonstrates a  method  for  patterning  the
ferrimagnetic   coordination  compound   vanadium  tetracyanoethylene.
Standard electron-beam lithography of PMMA/P(MMA-MAA) bilayers is used
in conjunction with pre-growth  aluminum encapsulation and post-growth
dichloromethane  liftoff   to  pattern  \VTCNE  thin   films  with  no
degradation  of the  microwave magnetic  properties. The  sidewalls of
structures  patterned  in  this  way  are  sloped,  allowing  for  the
investigation and  quantitative modeling  of spin-wave  confinement in
magnetic structures  with soft boundary conditions.   Patterned \VTCNE
films with features down to \SI{25}{\um}  exhibit a high \(Q\) of over
3,700  and  ultra-low damping  of  \num{4.60  \pm 0.44e-5}  which  are
competitive with unpatterned  YIG and lower than  all existing reports
of  patterned  YIG microstructures.\cite{Hahn_2014,  Jungfleisch_2015,
    Krysztofik_2017,  Li_2016,   Zhu_2017}  The  versatility   of  the
patterning and  deposition conditions of \VTCNEs,  in combination with
its  ultra-low  magnetic  damping,  position  \VTCNE  as  a  promising
candidate  for   incorporation  into  magnetoelectric   devices  where
low-loss,  highly coherent,  magnon  excitation  are desirable.   Such
applications   range   from   microwave  communications   to   quantum
information.

\section*{Supplementary Materials}
See supplementary  material for the  a detailed description  of sample
fabrication,   measurement  techniques,   simulations,  and   analytic
calculations.

\begin{acknowledgments}
    This  work is  supported  by Emerging  Frontiers  in Research  and
    Innovation (EFRI) Grant No.  EFMA-1741666. The authors acknowledge
    the NanoSystems Laboratory at The Ohio State University.
\end{acknowledgments}

\nocite{*}
\bibliography{Paperbib}

\begin{thebibliography}{53}%
\makeatletter
\providecommand \@ifxundefined [1]{%
 \@ifx{#1\undefined}
}%
\providecommand \@ifnum [1]{%
 \ifnum #1\expandafter \@firstoftwo
 \else \expandafter \@secondoftwo
 \fi
}%
\providecommand \@ifx [1]{%
 \ifx #1\expandafter \@firstoftwo
 \else \expandafter \@secondoftwo
 \fi
}%
\providecommand \natexlab [1]{#1}%
\providecommand \enquote  [1]{``#1''}%
\providecommand \bibnamefont  [1]{#1}%
\providecommand \bibfnamefont [1]{#1}%
\providecommand \citenamefont [1]{#1}%
\providecommand \href@noop [0]{\@secondoftwo}%
\providecommand \href [0]{\begingroup \@sanitize@url \@href}%
\providecommand \@href[1]{\@@startlink{#1}\@@href}%
\providecommand \@@href[1]{\endgroup#1\@@endlink}%
\providecommand \@sanitize@url [0]{\catcode `\\12\catcode `\$12\catcode
  `\&12\catcode `\#12\catcode `\^12\catcode `\_12\catcode `\%12\relax}%
\providecommand \@@startlink[1]{}%
\providecommand \@@endlink[0]{}%
\providecommand \url  [0]{\begingroup\@sanitize@url \@url }%
\providecommand \@url [1]{\endgroup\@href {#1}{\urlprefix }}%
\providecommand \urlprefix  [0]{URL }%
\providecommand \Eprint [0]{\href }%
\providecommand \doibase [0]{http://dx.doi.org/}%
\providecommand \selectlanguage [0]{\@gobble}%
\providecommand \bibinfo  [0]{\@secondoftwo}%
\providecommand \bibfield  [0]{\@secondoftwo}%
\providecommand \translation [1]{[#1]}%
\providecommand \BibitemOpen [0]{}%
\providecommand \bibitemStop [0]{}%
\providecommand \bibitemNoStop [0]{.\EOS\space}%
\providecommand \EOS [0]{\spacefactor3000\relax}%
\providecommand \BibitemShut  [1]{\csname bibitem#1\endcsname}%
\let\auto@bib@innerbib\@empty
\bibitem [{\citenamefont {Cheng}\ \emph {et~al.}(2018)\citenamefont {Cheng},
  \citenamefont {Peng}, \citenamefont {Hu}, \citenamefont {Zhou},\ and\
  \citenamefont {Liu}}]{Cheng_2018}%
  \BibitemOpen
  \bibfield  {author} {\bibinfo {author} {\bibfnamefont {Y.}~\bibnamefont
  {Cheng}}, \bibinfo {author} {\bibfnamefont {B.}~\bibnamefont {Peng}},
  \bibinfo {author} {\bibfnamefont {Z.}~\bibnamefont {Hu}}, \bibinfo {author}
  {\bibfnamefont {Z.}~\bibnamefont {Zhou}}, \ and\ \bibinfo {author}
  {\bibfnamefont {M.}~\bibnamefont {Liu}},\ }\bibfield  {title} {\enquote
  {\bibinfo {title} {Recent development and status of magnetoelectric materials
  and devices},}\ }\href {\doibase 10.1016/j.physleta.2018.07.014} {\bibfield
  {journal} {\bibinfo  {journal} {Physics Letters A}\ }\textbf {\bibinfo
  {volume} {382}},\ \bibinfo {pages} {3018–3025} (\bibinfo {year}
  {2018})}\BibitemShut {NoStop}%
\bibitem [{\citenamefont {Nikitin}\ \emph {et~al.}(2015)\citenamefont
  {Nikitin}, \citenamefont {Ustinov}, \citenamefont {Semenov}, \citenamefont
  {Chumak}, \citenamefont {Serga}, \citenamefont {Vasyuchka}, \citenamefont
  {Lahderanta}, \citenamefont {Kalinikos},\ and\ \citenamefont
  {Hillebrands}}]{Nikitin_2015}%
  \BibitemOpen
  \bibfield  {author} {\bibinfo {author} {\bibfnamefont {A.}~\bibnamefont
  {Nikitin}}, \bibinfo {author} {\bibfnamefont {A.}~\bibnamefont {Ustinov}},
  \bibinfo {author} {\bibfnamefont {A.}~\bibnamefont {Semenov}}, \bibinfo
  {author} {\bibfnamefont {A.}~\bibnamefont {Chumak}}, \bibinfo {author}
  {\bibfnamefont {A.}~\bibnamefont {Serga}}, \bibinfo {author} {\bibfnamefont
  {V.~I.}\ \bibnamefont {Vasyuchka}}, \bibinfo {author} {\bibfnamefont
  {E.}~\bibnamefont {Lahderanta}}, \bibinfo {author} {\bibfnamefont
  {B.}~\bibnamefont {Kalinikos}}, \ and\ \bibinfo {author} {\bibfnamefont
  {B.}~\bibnamefont {Hillebrands}},\ }\bibfield  {title} {\enquote {\bibinfo
  {title} {Width-modulated magnonic crystal and its application for spin-wave
  logic},}\ }\href {\doibase 10.1109/eumc.2015.7345998} {\bibfield  {journal}
  {\bibinfo  {journal} {2015 European Microwave Conference (EuMC)}\ } (\bibinfo
  {year} {2015}),\ 10.1109/eumc.2015.7345998}\BibitemShut {NoStop}%
\bibitem [{\citenamefont {Chumak}, \citenamefont {Serga},\ and\ \citenamefont
  {Hillebrands}(2017)}]{Chumak_2017}%
  \BibitemOpen
  \bibfield  {author} {\bibinfo {author} {\bibfnamefont {A.~V.}\ \bibnamefont
  {Chumak}}, \bibinfo {author} {\bibfnamefont {A.~A.}\ \bibnamefont {Serga}}, \
  and\ \bibinfo {author} {\bibfnamefont {B.}~\bibnamefont {Hillebrands}},\
  }\bibfield  {title} {\enquote {\bibinfo {title} {Magnonic crystals for data
  processing},}\ }\href {\doibase 10.1088/1361-6463/aa6a65} {\bibfield
  {journal} {\bibinfo  {journal} {Journal of Physics D: Applied Physics}\
  }\textbf {\bibinfo {volume} {50}},\ \bibinfo {pages} {244001} (\bibinfo
  {year} {2017})}\BibitemShut {NoStop}%
\bibitem [{\citenamefont {Chang}\ \emph {et~al.}(2014)\citenamefont {Chang},
  \citenamefont {Li}, \citenamefont {Zhang}, \citenamefont {Liu}, \citenamefont
  {Hoffmann}, \citenamefont {Deng},\ and\ \citenamefont
  {Wu}}]{Houchen_Chang_2014}%
  \BibitemOpen
  \bibfield  {author} {\bibinfo {author} {\bibfnamefont {H.}~\bibnamefont
  {Chang}}, \bibinfo {author} {\bibfnamefont {P.}~\bibnamefont {Li}}, \bibinfo
  {author} {\bibfnamefont {W.}~\bibnamefont {Zhang}}, \bibinfo {author}
  {\bibfnamefont {T.}~\bibnamefont {Liu}}, \bibinfo {author} {\bibfnamefont
  {A.}~\bibnamefont {Hoffmann}}, \bibinfo {author} {\bibfnamefont
  {L.}~\bibnamefont {Deng}}, \ and\ \bibinfo {author} {\bibfnamefont
  {M.}~\bibnamefont {Wu}},\ }\bibfield  {title} {\enquote {\bibinfo {title}
  {Nanometer-thick yttrium iron garnet films with extremely low damping},}\
  }\href {\doibase 10.1109/lmag.2014.2350958} {\bibfield  {journal} {\bibinfo
  {journal} {IEEE Magnetics Letters}\ }\textbf {\bibinfo {volume} {5}},\
  \bibinfo {pages} {1–4} (\bibinfo {year} {2014})}\BibitemShut {NoStop}%
\bibitem [{\citenamefont {Hauser}\ \emph {et~al.}(2016)\citenamefont {Hauser},
  \citenamefont {Richter}, \citenamefont {Homonnay}, \citenamefont
  {Eisenschmidt}, \citenamefont {Qaid}, \citenamefont {Deniz}, \citenamefont
  {Hesse}, \citenamefont {Sawicki}, \citenamefont {Ebbinghaus},\ and\
  \citenamefont {Schmidt}}]{Hauser_2016}%
  \BibitemOpen
  \bibfield  {author} {\bibinfo {author} {\bibfnamefont {C.}~\bibnamefont
  {Hauser}}, \bibinfo {author} {\bibfnamefont {T.}~\bibnamefont {Richter}},
  \bibinfo {author} {\bibfnamefont {N.}~\bibnamefont {Homonnay}}, \bibinfo
  {author} {\bibfnamefont {C.}~\bibnamefont {Eisenschmidt}}, \bibinfo {author}
  {\bibfnamefont {M.}~\bibnamefont {Qaid}}, \bibinfo {author} {\bibfnamefont
  {H.}~\bibnamefont {Deniz}}, \bibinfo {author} {\bibfnamefont
  {D.}~\bibnamefont {Hesse}}, \bibinfo {author} {\bibfnamefont
  {M.}~\bibnamefont {Sawicki}}, \bibinfo {author} {\bibfnamefont {S.~G.}\
  \bibnamefont {Ebbinghaus}}, \ and\ \bibinfo {author} {\bibfnamefont
  {G.}~\bibnamefont {Schmidt}},\ }\bibfield  {title} {\enquote {\bibinfo
  {title} {Yttrium iron garnet thin films with very low damping obtained by
  recrystallization of amorphous material},}\ }\href {\doibase
  10.1038/srep20827} {\bibfield  {journal} {\bibinfo  {journal} {Scientific
  Reports}\ }\textbf {\bibinfo {volume} {6}} (\bibinfo {year} {2016}),\
  10.1038/srep20827}\BibitemShut {NoStop}%
\bibitem [{\citenamefont {Serga}, \citenamefont {Chumak},\ and\ \citenamefont
  {Hillebrands}(2010)}]{Serga_2010}%
  \BibitemOpen
  \bibfield  {author} {\bibinfo {author} {\bibfnamefont {A.~A.}\ \bibnamefont
  {Serga}}, \bibinfo {author} {\bibfnamefont {A.~V.}\ \bibnamefont {Chumak}}, \
  and\ \bibinfo {author} {\bibfnamefont {B.}~\bibnamefont {Hillebrands}},\
  }\bibfield  {title} {\enquote {\bibinfo {title} {{YIG magnonics}},}\ }\href
  {\doibase 10.1088/0022-3727/43/26/264002} {\bibfield  {journal} {\bibinfo
  {journal} {Journal of Physics D: Applied Physics}\ }\textbf {\bibinfo
  {volume} {43}},\ \bibinfo {pages} {264002} (\bibinfo {year}
  {2010})}\BibitemShut {NoStop}%
\bibitem [{\citenamefont {Schneider}\ \emph {et~al.}(2008)\citenamefont
  {Schneider}, \citenamefont {Serga}, \citenamefont {Leven}, \citenamefont
  {Hillebrands}, \citenamefont {Stamps},\ and\ \citenamefont
  {Kostylev}}]{Schneider_2008}%
  \BibitemOpen
  \bibfield  {author} {\bibinfo {author} {\bibfnamefont {T.}~\bibnamefont
  {Schneider}}, \bibinfo {author} {\bibfnamefont {A.~A.}\ \bibnamefont
  {Serga}}, \bibinfo {author} {\bibfnamefont {B.}~\bibnamefont {Leven}},
  \bibinfo {author} {\bibfnamefont {B.}~\bibnamefont {Hillebrands}}, \bibinfo
  {author} {\bibfnamefont {R.~L.}\ \bibnamefont {Stamps}}, \ and\ \bibinfo
  {author} {\bibfnamefont {M.~P.}\ \bibnamefont {Kostylev}},\ }\bibfield
  {title} {\enquote {\bibinfo {title} {Realization of spin-wave logic gates},}\
  }\href {\doibase 10.1063/1.2834714} {\bibfield  {journal} {\bibinfo
  {journal} {Applied Physics Letters}\ }\textbf {\bibinfo {volume} {92}},\
  \bibinfo {pages} {022505} (\bibinfo {year} {2008})}\BibitemShut {NoStop}%
\bibitem [{\citenamefont {Zahwe}\ \emph {et~al.}(2010)\citenamefont {Zahwe},
  \citenamefont {Abdel~Samad}, \citenamefont {Sauviac}, \citenamefont
  {Chatelon}, \citenamefont {Blanc~Mignon}, \citenamefont {Rousseau},
  \citenamefont {Le~Berre},\ and\ \citenamefont {Givord}}]{Zahwe_2010}%
  \BibitemOpen
  \bibfield  {author} {\bibinfo {author} {\bibfnamefont {O.}~\bibnamefont
  {Zahwe}}, \bibinfo {author} {\bibfnamefont {B.}~\bibnamefont {Abdel~Samad}},
  \bibinfo {author} {\bibfnamefont {B.}~\bibnamefont {Sauviac}}, \bibinfo
  {author} {\bibfnamefont {J.~P.}\ \bibnamefont {Chatelon}}, \bibinfo {author}
  {\bibfnamefont {M.~F.}\ \bibnamefont {Blanc~Mignon}}, \bibinfo {author}
  {\bibfnamefont {J.~J.}\ \bibnamefont {Rousseau}}, \bibinfo {author}
  {\bibfnamefont {M.}~\bibnamefont {Le~Berre}}, \ and\ \bibinfo {author}
  {\bibfnamefont {D.}~\bibnamefont {Givord}},\ }\bibfield  {title} {\enquote
  {\bibinfo {title} {Yig thin film used to miniaturize a coplanar junction
  circulator},}\ }\href {\doibase 10.1163/156939310790322073} {\bibfield
  {journal} {\bibinfo  {journal} {Journal of Electromagnetic Waves and
  Applications}\ }\textbf {\bibinfo {volume} {24}},\ \bibinfo {pages} {25–32}
  (\bibinfo {year} {2010})}\BibitemShut {NoStop}%
\bibitem [{\citenamefont {Chumak}(2019)}]{Chumak_2019}%
  \BibitemOpen
  \bibfield  {author} {\bibinfo {author} {\bibfnamefont {A.~V.}\ \bibnamefont
  {Chumak}},\ }\bibfield  {title} {\enquote {\bibinfo {title} {Magnon
  spintronics},}\ }\href {\doibase 10.1201/9780429423079-6} {\bibfield
  {journal} {\bibinfo  {journal} {{Spintronics handbook: Spin transport and
  magnetism, second edition}}\ ,\ \bibinfo {pages} {247–302}} (\bibinfo
  {year} {2019})}\BibitemShut {NoStop}%
\bibitem [{\citenamefont {Hahn}\ \emph {et~al.}(2014)\citenamefont {Hahn},
  \citenamefont {Naletov}, \citenamefont {de~Loubens}, \citenamefont {Klein},
  \citenamefont {d'~Allivy~Kelly}, \citenamefont {Anane}, \citenamefont
  {Bernard}, \citenamefont {Jacquet}, \citenamefont {Bortolotti}, \citenamefont
  {Cros},\ and\ \citenamefont {et~al.}}]{Hahn_2014}%
  \BibitemOpen
  \bibfield  {author} {\bibinfo {author} {\bibfnamefont {C.}~\bibnamefont
  {Hahn}}, \bibinfo {author} {\bibfnamefont {V.~V.}\ \bibnamefont {Naletov}},
  \bibinfo {author} {\bibfnamefont {G.}~\bibnamefont {de~Loubens}}, \bibinfo
  {author} {\bibfnamefont {O.}~\bibnamefont {Klein}}, \bibinfo {author}
  {\bibfnamefont {O.}~\bibnamefont {d'~Allivy~Kelly}}, \bibinfo {author}
  {\bibfnamefont {A.}~\bibnamefont {Anane}}, \bibinfo {author} {\bibfnamefont
  {R.}~\bibnamefont {Bernard}}, \bibinfo {author} {\bibfnamefont
  {E.}~\bibnamefont {Jacquet}}, \bibinfo {author} {\bibfnamefont
  {P.}~\bibnamefont {Bortolotti}}, \bibinfo {author} {\bibfnamefont
  {V.}~\bibnamefont {Cros}}, \ and\ \bibinfo {author} {\bibnamefont {et~al.}},\
  }\bibfield  {title} {\enquote {\bibinfo {title} {{Measurement of the
  intrinsic damping constant in individual nanodisks of {\symb{Y_3Fe_5O_{12}}}
  and {\symb{Y_3Fe_5O_{12}}}|Pt}},}\ }\href {\doibase 10.1063/1.4871516}
  {\bibfield  {journal} {\bibinfo  {journal} {Applied Physics Letters}\
  }\textbf {\bibinfo {volume} {104}},\ \bibinfo {pages} {152410} (\bibinfo
  {year} {2014})}\BibitemShut {NoStop}%
\bibitem [{\citenamefont {Jungfleisch}\ \emph {et~al.}(2015)\citenamefont
  {Jungfleisch}, \citenamefont {Zhang}, \citenamefont {Jiang}, \citenamefont
  {Chang}, \citenamefont {Sklenar}, \citenamefont {Wu}, \citenamefont
  {Pearson}, \citenamefont {Bhattacharya}, \citenamefont {Ketterson},
  \citenamefont {Wu},\ and\ \citenamefont {et~al.}}]{Jungfleisch_2015}%
  \BibitemOpen
  \bibfield  {author} {\bibinfo {author} {\bibfnamefont {M.~B.}\ \bibnamefont
  {Jungfleisch}}, \bibinfo {author} {\bibfnamefont {W.}~\bibnamefont {Zhang}},
  \bibinfo {author} {\bibfnamefont {W.}~\bibnamefont {Jiang}}, \bibinfo
  {author} {\bibfnamefont {H.}~\bibnamefont {Chang}}, \bibinfo {author}
  {\bibfnamefont {J.}~\bibnamefont {Sklenar}}, \bibinfo {author} {\bibfnamefont
  {S.~M.}\ \bibnamefont {Wu}}, \bibinfo {author} {\bibfnamefont {J.~E.}\
  \bibnamefont {Pearson}}, \bibinfo {author} {\bibfnamefont {A.}~\bibnamefont
  {Bhattacharya}}, \bibinfo {author} {\bibfnamefont {J.~B.}\ \bibnamefont
  {Ketterson}}, \bibinfo {author} {\bibfnamefont {M.}~\bibnamefont {Wu}}, \
  and\ \bibinfo {author} {\bibnamefont {et~al.}},\ }\bibfield  {title}
  {\enquote {\bibinfo {title} {Spin waves in micro-structured yttrium iron
  garnet nanometer-thick films},}\ }\href {\doibase 10.1063/1.4916027}
  {\bibfield  {journal} {\bibinfo  {journal} {Journal of Applied Physics}\
  }\textbf {\bibinfo {volume} {117}},\ \bibinfo {pages} {17D128} (\bibinfo
  {year} {2015})}\BibitemShut {NoStop}%
\bibitem [{\citenamefont {Krysztofik}\ \emph {et~al.}(2017)\citenamefont
  {Krysztofik}, \citenamefont {Coy}, \citenamefont {Kuświk}, \citenamefont
  {Załęski}, \citenamefont {Głowiński},\ and\ \citenamefont
  {Dubowik}}]{Krysztofik_2017}%
  \BibitemOpen
  \bibfield  {author} {\bibinfo {author} {\bibfnamefont {A.}~\bibnamefont
  {Krysztofik}}, \bibinfo {author} {\bibfnamefont {L.~E.}\ \bibnamefont {Coy}},
  \bibinfo {author} {\bibfnamefont {P.}~\bibnamefont {Kuświk}}, \bibinfo
  {author} {\bibfnamefont {K.}~\bibnamefont {Załęski}}, \bibinfo {author}
  {\bibfnamefont {H.}~\bibnamefont {Głowiński}}, \ and\ \bibinfo {author}
  {\bibfnamefont {J.}~\bibnamefont {Dubowik}},\ }\bibfield  {title} {\enquote
  {\bibinfo {title} {Ultra-low damping in lift-off structured yttrium iron
  garnet thin films},}\ }\href {\doibase 10.1063/1.5002004} {\bibfield
  {journal} {\bibinfo  {journal} {Applied Physics Letters}\ }\textbf {\bibinfo
  {volume} {111}},\ \bibinfo {pages} {192404} (\bibinfo {year}
  {2017})}\BibitemShut {NoStop}%
\bibitem [{\citenamefont {Li}\ \emph {et~al.}(2016)\citenamefont {Li},
  \citenamefont {Zhang}, \citenamefont {Ding}, \citenamefont {Pearson},
  \citenamefont {Novosad},\ and\ \citenamefont {Hoffmann}}]{Li_2016}%
  \BibitemOpen
  \bibfield  {author} {\bibinfo {author} {\bibfnamefont {S.}~\bibnamefont
  {Li}}, \bibinfo {author} {\bibfnamefont {W.}~\bibnamefont {Zhang}}, \bibinfo
  {author} {\bibfnamefont {J.}~\bibnamefont {Ding}}, \bibinfo {author}
  {\bibfnamefont {J.~E.}\ \bibnamefont {Pearson}}, \bibinfo {author}
  {\bibfnamefont {V.}~\bibnamefont {Novosad}}, \ and\ \bibinfo {author}
  {\bibfnamefont {A.}~\bibnamefont {Hoffmann}},\ }\bibfield  {title} {\enquote
  {\bibinfo {title} {{Epitaxial patterning of nanometer-thick
  \symb{Y_3Fe_5O_{12}} films with low magnetic damping}},}\ }\href {\doibase
  10.1039/c5nr06808h} {\bibfield  {journal} {\bibinfo  {journal} {Nanoscale}\
  }\textbf {\bibinfo {volume} {8}},\ \bibinfo {pages} {388–394} (\bibinfo
  {year} {2016})}\BibitemShut {NoStop}%
\bibitem [{\citenamefont {Zhu}\ \emph {et~al.}(2017)\citenamefont {Zhu},
  \citenamefont {Chang}, \citenamefont {Franson}, \citenamefont {Liu},
  \citenamefont {Zhang}, \citenamefont {Johnston-Halperin}, \citenamefont
  {Wu},\ and\ \citenamefont {Tang}}]{Zhu_2017}%
  \BibitemOpen
  \bibfield  {author} {\bibinfo {author} {\bibfnamefont {N.}~\bibnamefont
  {Zhu}}, \bibinfo {author} {\bibfnamefont {H.}~\bibnamefont {Chang}}, \bibinfo
  {author} {\bibfnamefont {A.}~\bibnamefont {Franson}}, \bibinfo {author}
  {\bibfnamefont {T.}~\bibnamefont {Liu}}, \bibinfo {author} {\bibfnamefont
  {X.}~\bibnamefont {Zhang}}, \bibinfo {author} {\bibfnamefont
  {E.}~\bibnamefont {Johnston-Halperin}}, \bibinfo {author} {\bibfnamefont
  {M.}~\bibnamefont {Wu}}, \ and\ \bibinfo {author} {\bibfnamefont {H.~X.}\
  \bibnamefont {Tang}},\ }\bibfield  {title} {\enquote {\bibinfo {title}
  {{Patterned growth of crystalline \symb{Y_3Fe_5O_{12}} nanostructures with
  engineered magnetic shape anisotropy}},}\ }\href {\doibase 10.1063/1.4986474}
  {\bibfield  {journal} {\bibinfo  {journal} {Applied Physics Letters}\
  }\textbf {\bibinfo {volume} {110}},\ \bibinfo {pages} {252401} (\bibinfo
  {year} {2017})}\BibitemShut {NoStop}%
\bibitem [{\citenamefont {Onbasli}\ \emph {et~al.}(2014)\citenamefont
  {Onbasli}, \citenamefont {Kehlberger}, \citenamefont {Kim}, \citenamefont
  {Jakob}, \citenamefont {Kläui}, \citenamefont {Chumak}, \citenamefont
  {Hillebrands},\ and\ \citenamefont {Ross}}]{Onbasli_2014}%
  \BibitemOpen
  \bibfield  {author} {\bibinfo {author} {\bibfnamefont {M.~C.}\ \bibnamefont
  {Onbasli}}, \bibinfo {author} {\bibfnamefont {A.}~\bibnamefont {Kehlberger}},
  \bibinfo {author} {\bibfnamefont {D.~H.}\ \bibnamefont {Kim}}, \bibinfo
  {author} {\bibfnamefont {G.}~\bibnamefont {Jakob}}, \bibinfo {author}
  {\bibfnamefont {M.}~\bibnamefont {Kläui}}, \bibinfo {author} {\bibfnamefont
  {A.~V.}\ \bibnamefont {Chumak}}, \bibinfo {author} {\bibfnamefont
  {B.}~\bibnamefont {Hillebrands}}, \ and\ \bibinfo {author} {\bibfnamefont
  {C.~A.}\ \bibnamefont {Ross}},\ }\bibfield  {title} {\enquote {\bibinfo
  {title} {Pulsed laser deposition of epitaxial yttrium iron garnet films with
  low gilbert damping and bulk-like magnetization},}\ }\href {\doibase
  10.1063/1.4896936} {\bibfield  {journal} {\bibinfo  {journal} {APL
  Materials}\ }\textbf {\bibinfo {volume} {2}},\ \bibinfo {pages} {106102}
  (\bibinfo {year} {2014})}\BibitemShut {NoStop}%
\bibitem [{\citenamefont {Manuilov}\ and\ \citenamefont
  {Grishin}(2010)}]{Manuilov_2010}%
  \BibitemOpen
  \bibfield  {author} {\bibinfo {author} {\bibfnamefont {S.~A.}\ \bibnamefont
  {Manuilov}}\ and\ \bibinfo {author} {\bibfnamefont {A.~M.}\ \bibnamefont
  {Grishin}},\ }\bibfield  {title} {\enquote {\bibinfo {title} {{Pulsed laser
  deposited \symb{Y_3Fe_5O_{12}} films: Nature of magnetic anisotropy II}},}\
  }\href {\doibase 10.1063/1.3446840} {\bibfield  {journal} {\bibinfo
  {journal} {Journal of Applied Physics}\ }\textbf {\bibinfo {volume} {108}},\
  \bibinfo {pages} {013902} (\bibinfo {year} {2010})}\BibitemShut {NoStop}%
\bibitem [{\citenamefont {Manuilov}\ \emph {et~al.}(2009)\citenamefont
  {Manuilov}, \citenamefont {Fors}, \citenamefont {Khartsev},\ and\
  \citenamefont {Grishin}}]{Manuilov_2009}%
  \BibitemOpen
  \bibfield  {author} {\bibinfo {author} {\bibfnamefont {S.~A.}\ \bibnamefont
  {Manuilov}}, \bibinfo {author} {\bibfnamefont {R.}~\bibnamefont {Fors}},
  \bibinfo {author} {\bibfnamefont {S.~I.}\ \bibnamefont {Khartsev}}, \ and\
  \bibinfo {author} {\bibfnamefont {A.~M.}\ \bibnamefont {Grishin}},\
  }\bibfield  {title} {\enquote {\bibinfo {title} {{Submicron
  \symb{Y_3Fe_5O_{12}} film magnetostatic wave band pass filters}},}\ }\href
  {\doibase 10.1063/1.3075816} {\bibfield  {journal} {\bibinfo  {journal}
  {Journal of Applied Physics}\ }\textbf {\bibinfo {volume} {105}},\ \bibinfo
  {pages} {033917} (\bibinfo {year} {2009})}\BibitemShut {NoStop}%
\bibitem [{\citenamefont {Harberts}\ \emph {et~al.}(2015)\citenamefont
  {Harberts}, \citenamefont {Lu}, \citenamefont {Yu}, \citenamefont {Epstein},\
  and\ \citenamefont {Johnston-Halperin}}]{Harberts_2015}%
  \BibitemOpen
  \bibfield  {author} {\bibinfo {author} {\bibfnamefont {M.}~\bibnamefont
  {Harberts}}, \bibinfo {author} {\bibfnamefont {Y.}~\bibnamefont {Lu}},
  \bibinfo {author} {\bibfnamefont {H.}~\bibnamefont {Yu}}, \bibinfo {author}
  {\bibfnamefont {A.~J.}\ \bibnamefont {Epstein}}, \ and\ \bibinfo {author}
  {\bibfnamefont {E.}~\bibnamefont {Johnston-Halperin}},\ }\bibfield  {title}
  {\enquote {\bibinfo {title} {Chemical vapor deposition of an organic magnet,
  vanadium tetracyanoethylene},}\ }\href {\doibase 10.3791/52891} {\bibfield
  {journal} {\bibinfo  {journal} {Journal of Visualized Experiments}\ }
  (\bibinfo {year} {2015}),\ 10.3791/52891}\BibitemShut {NoStop}%
\bibitem [{\citenamefont {Yu}\ \emph {et~al.}(2014)\citenamefont {Yu},
  \citenamefont {Harberts}, \citenamefont {Adur}, \citenamefont {Lu},
  \citenamefont {Hammel}, \citenamefont {Johnston-Halperin},\ and\
  \citenamefont {Epstein}}]{Yu_2014}%
  \BibitemOpen
  \bibfield  {author} {\bibinfo {author} {\bibfnamefont {H.}~\bibnamefont
  {Yu}}, \bibinfo {author} {\bibfnamefont {M.}~\bibnamefont {Harberts}},
  \bibinfo {author} {\bibfnamefont {R.}~\bibnamefont {Adur}}, \bibinfo {author}
  {\bibfnamefont {Y.}~\bibnamefont {Lu}}, \bibinfo {author} {\bibfnamefont
  {P.~C.}\ \bibnamefont {Hammel}}, \bibinfo {author} {\bibfnamefont
  {E.}~\bibnamefont {Johnston-Halperin}}, \ and\ \bibinfo {author}
  {\bibfnamefont {A.~J.}\ \bibnamefont {Epstein}},\ }\bibfield  {title}
  {\enquote {\bibinfo {title} {Ultra-narrow ferromagnetic resonance in
  organic-based thin films grown via low temperature chemical vapor
  deposition},}\ }\href {\doibase 10.1063/1.4887924} {\bibfield  {journal}
  {\bibinfo  {journal} {Applied Physics Letters}\ }\textbf {\bibinfo {volume}
  {105}},\ \bibinfo {pages} {012407} (\bibinfo {year} {2014})}\BibitemShut
  {NoStop}%
\bibitem [{\citenamefont {Zhu}\ \emph {et~al.}(2016)\citenamefont {Zhu},
  \citenamefont {Zhang}, \citenamefont {Froning}, \citenamefont {Flatté},
  \citenamefont {Johnston-Halperin},\ and\ \citenamefont {Tang}}]{Zhu_2016}%
  \BibitemOpen
  \bibfield  {author} {\bibinfo {author} {\bibfnamefont {N.}~\bibnamefont
  {Zhu}}, \bibinfo {author} {\bibfnamefont {X.}~\bibnamefont {Zhang}}, \bibinfo
  {author} {\bibfnamefont {I.~H.}\ \bibnamefont {Froning}}, \bibinfo {author}
  {\bibfnamefont {M.~E.}\ \bibnamefont {Flatté}}, \bibinfo {author}
  {\bibfnamefont {E.}~\bibnamefont {Johnston-Halperin}}, \ and\ \bibinfo
  {author} {\bibfnamefont {H.~X.}\ \bibnamefont {Tang}},\ }\bibfield  {title}
  {\enquote {\bibinfo {title} {Low loss spin wave resonances in organic-based
  ferrimagnet vanadium tetracyanoethylene thin films},}\ }\href {\doibase
  10.1063/1.4961579} {\bibfield  {journal} {\bibinfo  {journal} {Applied
  Physics Letters}\ }\textbf {\bibinfo {volume} {109}},\ \bibinfo {pages}
  {082402} (\bibinfo {year} {2016})}\BibitemShut {NoStop}%
\bibitem [{\citenamefont {Liu}\ \emph {et~al.}(2018)\citenamefont {Liu},
  \citenamefont {Zhang}, \citenamefont {Malissa}, \citenamefont {Groesbeck},
  \citenamefont {Kavand}, \citenamefont {McLaughlin}, \citenamefont {Jamali},
  \citenamefont {Hao}, \citenamefont {Sun}, \citenamefont {Davidson},\ and\
  \citenamefont {et~al.}}]{Liu_2018}%
  \BibitemOpen
  \bibfield  {author} {\bibinfo {author} {\bibfnamefont {H.}~\bibnamefont
  {Liu}}, \bibinfo {author} {\bibfnamefont {C.}~\bibnamefont {Zhang}}, \bibinfo
  {author} {\bibfnamefont {H.}~\bibnamefont {Malissa}}, \bibinfo {author}
  {\bibfnamefont {M.}~\bibnamefont {Groesbeck}}, \bibinfo {author}
  {\bibfnamefont {M.}~\bibnamefont {Kavand}}, \bibinfo {author} {\bibfnamefont
  {R.}~\bibnamefont {McLaughlin}}, \bibinfo {author} {\bibfnamefont
  {S.}~\bibnamefont {Jamali}}, \bibinfo {author} {\bibfnamefont
  {J.}~\bibnamefont {Hao}}, \bibinfo {author} {\bibfnamefont {D.}~\bibnamefont
  {Sun}}, \bibinfo {author} {\bibfnamefont {R.~A.}\ \bibnamefont {Davidson}}, \
  and\ \bibinfo {author} {\bibnamefont {et~al.}},\ }\bibfield  {title}
  {\enquote {\bibinfo {title} {Organic-based magnon spintronics},}\ }\href
  {\doibase 10.1038/s41563-018-0035-3} {\bibfield  {journal} {\bibinfo
  {journal} {Nature Materials}\ }\textbf {\bibinfo {volume} {17}},\ \bibinfo
  {pages} {308–312} (\bibinfo {year} {2018})}\BibitemShut {NoStop}%
\bibitem [{\citenamefont {Chilcote}\ \emph {et~al.}(2019)\citenamefont
  {Chilcote}, \citenamefont {Harberts}, \citenamefont {Fuhrman}, \citenamefont
  {Lehmann}, \citenamefont {Lu}, \citenamefont {Franson}, \citenamefont {Yu},
  \citenamefont {Zhu}, \citenamefont {Tang}, \citenamefont {Schmidt},\ and\
  \citenamefont {Johnston-Halperin}}]{chilcote19:spin}%
  \BibitemOpen
  \bibfield  {author} {\bibinfo {author} {\bibfnamefont {M.}~\bibnamefont
  {Chilcote}}, \bibinfo {author} {\bibfnamefont {M.}~\bibnamefont {Harberts}},
  \bibinfo {author} {\bibfnamefont {B.}~\bibnamefont {Fuhrman}}, \bibinfo
  {author} {\bibfnamefont {K.}~\bibnamefont {Lehmann}}, \bibinfo {author}
  {\bibfnamefont {Y.}~\bibnamefont {Lu}}, \bibinfo {author} {\bibfnamefont
  {A.}~\bibnamefont {Franson}}, \bibinfo {author} {\bibfnamefont
  {H.}~\bibnamefont {Yu}}, \bibinfo {author} {\bibfnamefont {N.}~\bibnamefont
  {Zhu}}, \bibinfo {author} {\bibfnamefont {H.}~\bibnamefont {Tang}}, \bibinfo
  {author} {\bibfnamefont {G.}~\bibnamefont {Schmidt}}, \ and\ \bibinfo
  {author} {\bibfnamefont {E.}~\bibnamefont {Johnston-Halperin}},\ }\href@noop
  {} {\enquote {\bibinfo {title} {Spin-wave confinement and coupling in
  organic-based magnetic nanostructures},}\ } (\bibinfo {year} {2019}),\
  \Eprint {http://arxiv.org/abs/1901.03286v1} {arXiv:1901.03286v1
  [cond-mat.mtrl-sci]} \BibitemShut {NoStop}%
\bibitem [{\citenamefont {Froning}\ \emph {et~al.}(2015)\citenamefont
  {Froning}, \citenamefont {Harberts}, \citenamefont {Lu}, \citenamefont {Yu},
  \citenamefont {Epstein},\ and\ \citenamefont
  {Johnston-Halperin}}]{Froning_2015}%
  \BibitemOpen
  \bibfield  {author} {\bibinfo {author} {\bibfnamefont {I.~H.}\ \bibnamefont
  {Froning}}, \bibinfo {author} {\bibfnamefont {M.}~\bibnamefont {Harberts}},
  \bibinfo {author} {\bibfnamefont {Y.}~\bibnamefont {Lu}}, \bibinfo {author}
  {\bibfnamefont {H.}~\bibnamefont {Yu}}, \bibinfo {author} {\bibfnamefont
  {A.~J.}\ \bibnamefont {Epstein}}, \ and\ \bibinfo {author} {\bibfnamefont
  {E.}~\bibnamefont {Johnston-Halperin}},\ }\bibfield  {title} {\enquote
  {\bibinfo {title} {Thin-film encapsulation of the air-sensitive organic-based
  ferrimagnet vanadium tetracyanoethylene},}\ }\href {\doibase
  10.1063/1.4916241} {\bibfield  {journal} {\bibinfo  {journal} {Applied
  Physics Letters}\ }\textbf {\bibinfo {volume} {106}},\ \bibinfo {pages}
  {122403} (\bibinfo {year} {2015})}\BibitemShut {NoStop}%
\bibitem [{\citenamefont {Bardoliwalla}()}]{Bardoliwalla}%
  \BibitemOpen
  \bibfield  {author} {\bibinfo {author} {\bibfnamefont {D.}~\bibnamefont
  {Bardoliwalla}},\ }\bibfield  {title} {\enquote {\bibinfo {title} {Fast
  curing, low exotherm epoxy potting and encapsulating systems},}\ }\href
  {\doibase 10.1109/eeic.1997.651053} {\bibfield  {journal} {\bibinfo
  {journal} {{Proceedings: Electrical insulation conference and electrical
  manufacturing and coil winding conference}}\
  }10.1109/eeic.1997.651053}\BibitemShut {NoStop}%
\bibitem [{\citenamefont {Thorum}, \citenamefont {Pokhodnya},\ and\
  \citenamefont {Miller}(2006)}]{Thorum_2006}%
  \BibitemOpen
  \bibfield  {author} {\bibinfo {author} {\bibfnamefont {M.~S.}\ \bibnamefont
  {Thorum}}, \bibinfo {author} {\bibfnamefont {K.~I.}\ \bibnamefont
  {Pokhodnya}}, \ and\ \bibinfo {author} {\bibfnamefont {J.~S.}\ \bibnamefont
  {Miller}},\ }\bibfield  {title} {\enquote {\bibinfo {title} {{Solvent
  enhancement of the magnetic ordering temperature (\symb{T_c}) of the room
  temperature \VTCNE $\cdot$S (S=solvent, TCNE=tetracyanoethylene;
  $x$$\approx$2) magnet}},}\ }\href {\doibase 10.1016/j.poly.2005.12.007}
  {\bibfield  {journal} {\bibinfo  {journal} {Polyhedron}\ }\textbf {\bibinfo
  {volume} {25}},\ \bibinfo {pages} {1927–1930} (\bibinfo {year}
  {2006})}\BibitemShut {NoStop}%
\bibitem [{\citenamefont {Pokhodnya}\ \emph {et~al.}(2001)\citenamefont
  {Pokhodnya}, \citenamefont {Pejakovic}, \citenamefont {Epstein},\ and\
  \citenamefont {Miller}}]{Pokhodnya_2001}%
  \BibitemOpen
  \bibfield  {author} {\bibinfo {author} {\bibfnamefont {K.~I.}\ \bibnamefont
  {Pokhodnya}}, \bibinfo {author} {\bibfnamefont {D.}~\bibnamefont
  {Pejakovic}}, \bibinfo {author} {\bibfnamefont {A.~J.}\ \bibnamefont
  {Epstein}}, \ and\ \bibinfo {author} {\bibfnamefont {J.~S.}\ \bibnamefont
  {Miller}},\ }\bibfield  {title} {\enquote {\bibinfo {title} {{Effect of
  solvent on the magnetic properties of the high-temperature \VTCNE
  molecule-based magnet}},}\ }\href {\doibase 10.1103/physrevb.63.174408}
  {\bibfield  {journal} {\bibinfo  {journal} {Physical Review B}\ }\textbf
  {\bibinfo {volume} {63}} (\bibinfo {year} {2001}),\
  10.1103/physrevb.63.174408}\BibitemShut {NoStop}%
\bibitem [{\citenamefont {Zhang}\ \emph {et~al.}(1996)\citenamefont {Zhang},
  \citenamefont {Miller}, \citenamefont {Vazquez}, \citenamefont {Zhou},
  \citenamefont {Brinckerhoff},\ and\ \citenamefont {Epstein}}]{Zhang_1996}%
  \BibitemOpen
  \bibfield  {author} {\bibinfo {author} {\bibfnamefont {J.}~\bibnamefont
  {Zhang}}, \bibinfo {author} {\bibfnamefont {J.~S.}\ \bibnamefont {Miller}},
  \bibinfo {author} {\bibfnamefont {C.}~\bibnamefont {Vazquez}}, \bibinfo
  {author} {\bibfnamefont {P.}~\bibnamefont {Zhou}}, \bibinfo {author}
  {\bibfnamefont {W.~B.}\ \bibnamefont {Brinckerhoff}}, \ and\ \bibinfo
  {author} {\bibfnamefont {A.~J.}\ \bibnamefont {Epstein}},\ }\bibfield
  {title} {\enquote {\bibinfo {title} {{Improved synthesis of the
  \ensuremath{\mathrm{V(tetracyanoethylene)}_x} $\cdot$ y(solvent)
  room-temperature magnet: Doubling of the magnetization at room
  temperature}},}\ }\href {\doibase 10.1021/bk-1996-0644.ch021} {\bibfield
  {journal} {\bibinfo  {journal} {Molecule-Based Magnetic Materials}\ ,\
  \bibinfo {pages} {311–318}} (\bibinfo {year} {1996})}\BibitemShut {NoStop}%
\bibitem [{\citenamefont {Tseng}\ \emph {et~al.}(2003)\citenamefont {Tseng},
  \citenamefont {Chen}, \citenamefont {Chen},\ and\ \citenamefont
  {Ma}}]{Tseng_2003}%
  \BibitemOpen
  \bibfield  {author} {\bibinfo {author} {\bibfnamefont {A.}~\bibnamefont
  {Tseng}}, \bibinfo {author} {\bibfnamefont {K.}~\bibnamefont {Chen}},
  \bibinfo {author} {\bibfnamefont {C.}~\bibnamefont {Chen}}, \ and\ \bibinfo
  {author} {\bibfnamefont {K.}~\bibnamefont {Ma}},\ }\bibfield  {title}
  {\enquote {\bibinfo {title} {Electron beam lithography in nanoscale
  fabrication: Recent development},}\ }\href {\doibase
  10.1109/tepm.2003.817714} {\bibfield  {journal} {\bibinfo  {journal} {IEEE
  Transactions on Electronics Packaging Manufacturing}\ }\textbf {\bibinfo
  {volume} {26}},\ \bibinfo {pages} {141–149} (\bibinfo {year}
  {2003})}\BibitemShut {NoStop}%
\bibitem [{\citenamefont {Tennant}\ and\ \citenamefont
  {Bleier}(2016)}]{Tennant_2016}%
  \BibitemOpen
  \bibfield  {author} {\bibinfo {author} {\bibfnamefont {D.}~\bibnamefont
  {Tennant}}\ and\ \bibinfo {author} {\bibfnamefont {A.}~\bibnamefont
  {Bleier}},\ }\bibfield  {title} {\enquote {\bibinfo {title} {Electron beam
  lithography of nanostructures},}\ }\href {\doibase
  10.1016/b978-0-12-803581-8.09255-9} {\bibfield  {journal} {\bibinfo
  {journal} {Reference Module in Materials Science and Materials Engineering}\
  } (\bibinfo {year} {2016}),\ 10.1016/b978-0-12-803581-8.09255-9}\BibitemShut
  {NoStop}%
\bibitem [{\citenamefont {Hu}\ \emph {et~al.}(2014)\citenamefont {Hu},
  \citenamefont {Dyck}, \citenamefont {Chen}, \citenamefont {Hsiao},
  \citenamefont {Hu}, \citenamefont {Duscher}, \citenamefont {Dadmun},\ and\
  \citenamefont {Khomami}}]{Hu_2014}%
  \BibitemOpen
  \bibfield  {author} {\bibinfo {author} {\bibfnamefont {S.}~\bibnamefont
  {Hu}}, \bibinfo {author} {\bibfnamefont {O.}~\bibnamefont {Dyck}}, \bibinfo
  {author} {\bibfnamefont {H.}~\bibnamefont {Chen}}, \bibinfo {author}
  {\bibfnamefont {Y.-c.}\ \bibnamefont {Hsiao}}, \bibinfo {author}
  {\bibfnamefont {B.}~\bibnamefont {Hu}}, \bibinfo {author} {\bibfnamefont
  {G.}~\bibnamefont {Duscher}}, \bibinfo {author} {\bibfnamefont
  {M.}~\bibnamefont {Dadmun}}, \ and\ \bibinfo {author} {\bibfnamefont
  {B.}~\bibnamefont {Khomami}},\ }\bibfield  {title} {\enquote {\bibinfo
  {title} {The impact of selective solvents on the evolution of structure and
  function in solvent annealed organic photovoltaics},}\ }\href {\doibase
  10.1039/c4ra02257b} {\bibfield  {journal} {\bibinfo  {journal} {RSC Adv.}\
  }\textbf {\bibinfo {volume} {4}},\ \bibinfo {pages} {27931–27938} (\bibinfo
  {year} {2014})}\BibitemShut {NoStop}%
\bibitem [{\citenamefont {Tait}, \citenamefont {Smy},\ and\ \citenamefont
  {Brett}(1993)}]{Tait_1993}%
  \BibitemOpen
  \bibfield  {author} {\bibinfo {author} {\bibfnamefont {R.}~\bibnamefont
  {Tait}}, \bibinfo {author} {\bibfnamefont {T.}~\bibnamefont {Smy}}, \ and\
  \bibinfo {author} {\bibfnamefont {M.}~\bibnamefont {Brett}},\ }\bibfield
  {title} {\enquote {\bibinfo {title} {Modelling and characterization of
  columnar growth in evaporated films},}\ }\href {\doibase
  10.1016/0040-6090(93)90378-3} {\bibfield  {journal} {\bibinfo  {journal}
  {Thin Solid Films}\ }\textbf {\bibinfo {volume} {226}},\ \bibinfo {pages}
  {196–201} (\bibinfo {year} {1993})}\BibitemShut {NoStop}%
\bibitem [{\citenamefont {Shankar}\ and\ \citenamefont
  {Deshpande}(2000)}]{Shankar_2000}%
  \BibitemOpen
  \bibfield  {author} {\bibinfo {author} {\bibfnamefont {P.~N.}\ \bibnamefont
  {Shankar}}\ and\ \bibinfo {author} {\bibfnamefont {M.~D.}\ \bibnamefont
  {Deshpande}},\ }\bibfield  {title} {\enquote {\bibinfo {title} {Fluid
  mechanics in the driven cavity},}\ }\href {\doibase
  10.1146/annurev.fluid.32.1.93} {\bibfield  {journal} {\bibinfo  {journal}
  {Annual Review of Fluid Mechanics}\ }\textbf {\bibinfo {volume} {32}},\
  \bibinfo {pages} {93–136} (\bibinfo {year} {2000})}\BibitemShut {NoStop}%
\bibitem [{\citenamefont {Taneda}(1979)}]{Taneda_1979}%
  \BibitemOpen
  \bibfield  {author} {\bibinfo {author} {\bibfnamefont {S.}~\bibnamefont
  {Taneda}},\ }\bibfield  {title} {\enquote {\bibinfo {title} {Visualization of
  separating stokes flows},}\ }\href {\doibase 10.1143/jpsj.46.1935} {\bibfield
   {journal} {\bibinfo  {journal} {Journal of the Physical Society of Japan}\
  }\textbf {\bibinfo {volume} {46}},\ \bibinfo {pages} {1935–1942} (\bibinfo
  {year} {1979})}\BibitemShut {NoStop}%
\bibitem [{\citenamefont {Suhl}(1955)}]{Suhl_1955}%
  \BibitemOpen
  \bibfield  {author} {\bibinfo {author} {\bibfnamefont {H.}~\bibnamefont
  {Suhl}},\ }\bibfield  {title} {\enquote {\bibinfo {title} {Ferromagnetic
  resonance in nickel ferrite between one and two kilomegacycles},}\ }\href
  {\doibase 10.1103/physrev.97.555.2} {\bibfield  {journal} {\bibinfo
  {journal} {Physical Review}\ }\textbf {\bibinfo {volume} {97}},\ \bibinfo
  {pages} {555–557} (\bibinfo {year} {1955})}\BibitemShut {NoStop}%
\bibitem [{\citenamefont {Smit}\ and\ \citenamefont
  {Beljers}(1955)}]{Smit_1955}%
  \BibitemOpen
  \bibfield  {author} {\bibinfo {author} {\bibfnamefont {J.}~\bibnamefont
  {Smit}}\ and\ \bibinfo {author} {\bibfnamefont {H.~G.}\ \bibnamefont
  {Beljers}},\ }\bibfield  {title} {\enquote {\bibinfo {title} {{Ferromagnetic
  resonance absorption in \symb{BaFe_{12}O_{19}}, a highly anisotropic
  crystal}},}\ }\href@noop {} {\bibfield  {journal} {\bibinfo  {journal}
  {Philips Res. Rep.}\ }\textbf {\bibinfo {volume} {10}},\ \bibinfo {pages}
  {113--130} (\bibinfo {year} {1955})}\BibitemShut {NoStop}%
\bibitem [{\citenamefont {Baselgia}\ \emph {et~al.}(1988)\citenamefont
  {Baselgia}, \citenamefont {Warden}, \citenamefont {Waldner}, \citenamefont
  {Hutton}, \citenamefont {Drumheller}, \citenamefont {He}, \citenamefont
  {Wigen},\ and\ \citenamefont {Maryško}}]{Baselgia_1988}%
  \BibitemOpen
  \bibfield  {author} {\bibinfo {author} {\bibfnamefont {L.}~\bibnamefont
  {Baselgia}}, \bibinfo {author} {\bibfnamefont {M.}~\bibnamefont {Warden}},
  \bibinfo {author} {\bibfnamefont {F.}~\bibnamefont {Waldner}}, \bibinfo
  {author} {\bibfnamefont {S.~L.}\ \bibnamefont {Hutton}}, \bibinfo {author}
  {\bibfnamefont {J.~E.}\ \bibnamefont {Drumheller}}, \bibinfo {author}
  {\bibfnamefont {Y.~Q.}\ \bibnamefont {He}}, \bibinfo {author} {\bibfnamefont
  {P.~E.}\ \bibnamefont {Wigen}}, \ and\ \bibinfo {author} {\bibfnamefont
  {M.}~\bibnamefont {Maryško}},\ }\bibfield  {title} {\enquote {\bibinfo
  {title} {Derivation of the resonance frequency from the free energy of
  ferromagnets},}\ }\href {\doibase 10.1103/physrevb.38.2237} {\bibfield
  {journal} {\bibinfo  {journal} {Physical Review B}\ }\textbf {\bibinfo
  {volume} {38}},\ \bibinfo {pages} {2237–2242} (\bibinfo {year}
  {1988})}\BibitemShut {NoStop}%
\bibitem [{\citenamefont {Smit}\ and\ \citenamefont {Wijn}(1954)}]{Smit_1954}%
  \BibitemOpen
  \bibfield  {author} {\bibinfo {author} {\bibfnamefont {J.}~\bibnamefont
  {Smit}}\ and\ \bibinfo {author} {\bibfnamefont {H.}~\bibnamefont {Wijn}},\
  }\bibfield  {title} {\enquote {\bibinfo {title} {Physical properties of
  ferrites},}\ }\href {\doibase 10.1016/s0065-2539(08)60132-8} {\bibfield
  {journal} {\bibinfo  {journal} {Advances in Electronics and Electron
  Physics}\ ,\ \bibinfo {pages} {69–136}} (\bibinfo {year}
  {1954})}\BibitemShut {NoStop}%
\bibitem [{\citenamefont {Morrish}(2001)}]{Morrish_2001}%
  \BibitemOpen
  \bibfield  {author} {\bibinfo {author} {\bibfnamefont {A.~H.}\ \bibnamefont
  {Morrish}},\ }\bibfield  {title} {\enquote {\bibinfo {title} {The physical
  principles of magnetism},}\ }\href {\doibase 10.1109/9780470546581} {\
  (\bibinfo {year} {2001}),\ 10.1109/9780470546581}\BibitemShut {NoStop}%
\bibitem [{\citenamefont {Smith}\ \emph {et~al.}(2010)\citenamefont {Smith},
  \citenamefont {Nielsen}, \citenamefont {Christensen}, \citenamefont {Bahl},
  \citenamefont {Bjørk},\ and\ \citenamefont {Hattel}}]{Smith_2010}%
  \BibitemOpen
  \bibfield  {author} {\bibinfo {author} {\bibfnamefont {A.}~\bibnamefont
  {Smith}}, \bibinfo {author} {\bibfnamefont {K.~K.}\ \bibnamefont {Nielsen}},
  \bibinfo {author} {\bibfnamefont {D.~V.}\ \bibnamefont {Christensen}},
  \bibinfo {author} {\bibfnamefont {C.~R.~H.}\ \bibnamefont {Bahl}}, \bibinfo
  {author} {\bibfnamefont {R.}~\bibnamefont {Bjørk}}, \ and\ \bibinfo {author}
  {\bibfnamefont {J.}~\bibnamefont {Hattel}},\ }\bibfield  {title} {\enquote
  {\bibinfo {title} {The demagnetizing field of a nonuniform rectangular
  prism},}\ }\href {\doibase 10.1063/1.3385387} {\bibfield  {journal} {\bibinfo
   {journal} {Journal of Applied Physics}\ }\textbf {\bibinfo {volume} {107}},\
  \bibinfo {pages} {103910} (\bibinfo {year} {2010})}\BibitemShut {NoStop}%
\bibitem [{\citenamefont {Kraus}(1973)}]{Kraus_1973}%
  \BibitemOpen
  \bibfield  {author} {\bibinfo {author} {\bibfnamefont {L.}~\bibnamefont
  {Kraus}},\ }\bibfield  {title} {\enquote {\bibinfo {title} {The
  demagnetization tensor of a cylinder},}\ }\href {\doibase 10.1007/bf01593828}
  {\bibfield  {journal} {\bibinfo  {journal} {Czechoslovak Journal of Physics}\
  }\textbf {\bibinfo {volume} {23}},\ \bibinfo {pages} {512–519} (\bibinfo
  {year} {1973})}\BibitemShut {NoStop}%
\bibitem [{\citenamefont {Kalinikos}\ and\ \citenamefont
  {Slavin}(1986)}]{Kalinikos_1986}%
  \BibitemOpen
  \bibfield  {author} {\bibinfo {author} {\bibfnamefont {B.~A.}\ \bibnamefont
  {Kalinikos}}\ and\ \bibinfo {author} {\bibfnamefont {A.~N.}\ \bibnamefont
  {Slavin}},\ }\bibfield  {title} {\enquote {\bibinfo {title} {Theory of
  dipole-exchange spin wave spectrum for ferromagnetic films with mixed
  exchange boundary conditions},}\ }\href {\doibase
  10.1088/0022-3719/19/35/014} {\bibfield  {journal} {\bibinfo  {journal}
  {Journal of Physics C: Solid State Physics}\ }\textbf {\bibinfo {volume}
  {19}},\ \bibinfo {pages} {7013–7033} (\bibinfo {year} {1986})}\BibitemShut
  {NoStop}%
\bibitem [{\citenamefont {Skomski}\ \emph {et~al.}(2003)\citenamefont
  {Skomski}, \citenamefont {Kashyap}, \citenamefont {Qiang},\ and\
  \citenamefont {Sellmyer}}]{Skomski_2003}%
  \BibitemOpen
  \bibfield  {author} {\bibinfo {author} {\bibfnamefont {R.}~\bibnamefont
  {Skomski}}, \bibinfo {author} {\bibfnamefont {A.}~\bibnamefont {Kashyap}},
  \bibinfo {author} {\bibfnamefont {Y.}~\bibnamefont {Qiang}}, \ and\ \bibinfo
  {author} {\bibfnamefont {D.~J.}\ \bibnamefont {Sellmyer}},\ }\bibfield
  {title} {\enquote {\bibinfo {title} {Exchange through nonmagnetic insulating
  matrix},}\ }\href {\doibase 10.1063/1.1541633} {\bibfield  {journal}
  {\bibinfo  {journal} {Journal of Applied Physics}\ }\textbf {\bibinfo
  {volume} {93}},\ \bibinfo {pages} {6477–6479} (\bibinfo {year}
  {2003})}\BibitemShut {NoStop}%
\bibitem [{\citenamefont {Skomski}(2004)}]{Skomski_2004}%
  \BibitemOpen
  \bibfield  {author} {\bibinfo {author} {\bibfnamefont {R.}~\bibnamefont
  {Skomski}},\ }\bibfield  {title} {\enquote {\bibinfo {title} {Nanomagnetic
  scaling},}\ }\href {\doibase 10.1016/j.jmmm.2003.12.175} {\bibfield
  {journal} {\bibinfo  {journal} {Journal of Magnetism and Magnetic Materials}\
  }\textbf {\bibinfo {volume} {272-276}},\ \bibinfo {pages} {1476–1481}
  (\bibinfo {year} {2004})}\BibitemShut {NoStop}%
\bibitem [{\citenamefont {Moreno}\ \emph {et~al.}(2016)\citenamefont {Moreno},
  \citenamefont {Evans}, \citenamefont {Khmelevskyi}, \citenamefont {Muñoz},
  \citenamefont {Chantrell},\ and\ \citenamefont
  {Chubykalo-Fesenko}}]{Moreno_2016}%
  \BibitemOpen
  \bibfield  {author} {\bibinfo {author} {\bibfnamefont {R.}~\bibnamefont
  {Moreno}}, \bibinfo {author} {\bibfnamefont {R.~F.~L.}\ \bibnamefont
  {Evans}}, \bibinfo {author} {\bibfnamefont {S.}~\bibnamefont {Khmelevskyi}},
  \bibinfo {author} {\bibfnamefont {M.~C.}\ \bibnamefont {Muñoz}}, \bibinfo
  {author} {\bibfnamefont {R.~W.}\ \bibnamefont {Chantrell}}, \ and\ \bibinfo
  {author} {\bibfnamefont {O.}~\bibnamefont {Chubykalo-Fesenko}},\ }\bibfield
  {title} {\enquote {\bibinfo {title} {{Temperature-dependent exchange
  stiffness and domain wall width in Co}},}\ }\href {\doibase
  10.1103/physrevb.94.104433} {\bibfield  {journal} {\bibinfo  {journal}
  {Physical Review B}\ }\textbf {\bibinfo {volume} {94}} (\bibinfo {year}
  {2016}),\ 10.1103/physrevb.94.104433}\BibitemShut {NoStop}%
\bibitem [{\citenamefont {Vansteenkiste}\ \emph {et~al.}(2014)\citenamefont
  {Vansteenkiste}, \citenamefont {Leliaert}, \citenamefont {Dvornik},
  \citenamefont {Helsen}, \citenamefont {Garcia-Sanchez},\ and\ \citenamefont
  {Van~Waeyenberge}}]{Vansteenkiste_2014}%
  \BibitemOpen
  \bibfield  {author} {\bibinfo {author} {\bibfnamefont {A.}~\bibnamefont
  {Vansteenkiste}}, \bibinfo {author} {\bibfnamefont {J.}~\bibnamefont
  {Leliaert}}, \bibinfo {author} {\bibfnamefont {M.}~\bibnamefont {Dvornik}},
  \bibinfo {author} {\bibfnamefont {M.}~\bibnamefont {Helsen}}, \bibinfo
  {author} {\bibfnamefont {F.}~\bibnamefont {Garcia-Sanchez}}, \ and\ \bibinfo
  {author} {\bibfnamefont {B.}~\bibnamefont {Van~Waeyenberge}},\ }\bibfield
  {title} {\enquote {\bibinfo {title} {{The design and verification of
  MuMax3}},}\ }\href {\doibase 10.1063/1.4899186} {\bibfield  {journal}
  {\bibinfo  {journal} {AIP Advances}\ }\textbf {\bibinfo {volume} {4}},\
  \bibinfo {pages} {107133} (\bibinfo {year} {2014})}\BibitemShut {NoStop}%
\bibitem [{\citenamefont {Kalarickal}\ \emph {et~al.}(2006)\citenamefont
  {Kalarickal}, \citenamefont {Krivosik}, \citenamefont {Wu}, \citenamefont
  {Patton}, \citenamefont {Schneider}, \citenamefont {Kabos}, \citenamefont
  {Silva},\ and\ \citenamefont {Nibarger}}]{Kalarickal_2006}%
  \BibitemOpen
  \bibfield  {author} {\bibinfo {author} {\bibfnamefont {S.~S.}\ \bibnamefont
  {Kalarickal}}, \bibinfo {author} {\bibfnamefont {P.}~\bibnamefont
  {Krivosik}}, \bibinfo {author} {\bibfnamefont {M.}~\bibnamefont {Wu}},
  \bibinfo {author} {\bibfnamefont {C.~E.}\ \bibnamefont {Patton}}, \bibinfo
  {author} {\bibfnamefont {M.~L.}\ \bibnamefont {Schneider}}, \bibinfo {author}
  {\bibfnamefont {P.}~\bibnamefont {Kabos}}, \bibinfo {author} {\bibfnamefont
  {T.~J.}\ \bibnamefont {Silva}}, \ and\ \bibinfo {author} {\bibfnamefont
  {J.~P.}\ \bibnamefont {Nibarger}},\ }\bibfield  {title} {\enquote {\bibinfo
  {title} {Ferromagnetic resonance linewidth in metallic thin films: Comparison
  of measurement methods},}\ }\href {\doibase 10.1063/1.2197087} {\bibfield
  {journal} {\bibinfo  {journal} {Journal of Applied Physics}\ }\textbf
  {\bibinfo {volume} {99}},\ \bibinfo {pages} {093909} (\bibinfo {year}
  {2006})}\BibitemShut {NoStop}%
\bibitem [{\citenamefont {Heinrich}, \citenamefont {Cochran},\ and\
  \citenamefont {Hasegawa}(1985)}]{Heinrich_1985}%
  \BibitemOpen
  \bibfield  {author} {\bibinfo {author} {\bibfnamefont {B.}~\bibnamefont
  {Heinrich}}, \bibinfo {author} {\bibfnamefont {J.~F.}\ \bibnamefont
  {Cochran}}, \ and\ \bibinfo {author} {\bibfnamefont {R.}~\bibnamefont
  {Hasegawa}},\ }\bibfield  {title} {\enquote {\bibinfo {title} {Fmr
  linebroadening in metals due to two-magnon scattering},}\ }\href {\doibase
  10.1063/1.334991} {\bibfield  {journal} {\bibinfo  {journal} {Journal of
  Applied Physics}\ }\textbf {\bibinfo {volume} {57}},\ \bibinfo {pages}
  {3690–3692} (\bibinfo {year} {1985})}\BibitemShut {NoStop}%
\bibitem [{\citenamefont {Balinskiy}\ \emph {et~al.}(2017)\citenamefont
  {Balinskiy}, \citenamefont {Mongolov}, \citenamefont {Gutierrez},
  \citenamefont {Chiang}, \citenamefont {Slavin},\ and\ \citenamefont
  {Khitun}}]{Balinskiy_2017}%
  \BibitemOpen
  \bibfield  {author} {\bibinfo {author} {\bibfnamefont {M.}~\bibnamefont
  {Balinskiy}}, \bibinfo {author} {\bibfnamefont {B.}~\bibnamefont {Mongolov}},
  \bibinfo {author} {\bibfnamefont {D.}~\bibnamefont {Gutierrez}}, \bibinfo
  {author} {\bibfnamefont {H.}~\bibnamefont {Chiang}}, \bibinfo {author}
  {\bibfnamefont {A.}~\bibnamefont {Slavin}}, \ and\ \bibinfo {author}
  {\bibfnamefont {A.}~\bibnamefont {Khitun}},\ }\bibfield  {title} {\enquote
  {\bibinfo {title} {{Perpendicularly magnetized YIG-film resonators and
  waveguides with high operating power}},}\ }\href {\doibase 10.1063/1.4973497}
  {\bibfield  {journal} {\bibinfo  {journal} {AIP Advances}\ }\textbf {\bibinfo
  {volume} {7}},\ \bibinfo {pages} {056612} (\bibinfo {year}
  {2017})}\BibitemShut {NoStop}%
\bibitem [{\citenamefont {Chumak}, \citenamefont {Serga},\ and\ \citenamefont
  {Hillebrands}(2014)}]{Chumak_2014}%
  \BibitemOpen
  \bibfield  {author} {\bibinfo {author} {\bibfnamefont {A.~V.}\ \bibnamefont
  {Chumak}}, \bibinfo {author} {\bibfnamefont {A.~A.}\ \bibnamefont {Serga}}, \
  and\ \bibinfo {author} {\bibfnamefont {B.}~\bibnamefont {Hillebrands}},\
  }\bibfield  {title} {\enquote {\bibinfo {title} {Magnon transistor for
  all-magnon data processing},}\ }\href {\doibase 10.1038/ncomms5700}
  {\bibfield  {journal} {\bibinfo  {journal} {Nature Communications}\ }\textbf
  {\bibinfo {volume} {5}} (\bibinfo {year} {2014}),\
  10.1038/ncomms5700}\BibitemShut {NoStop}%
\bibitem [{\citenamefont {Cornelissen}\ \emph {et~al.}(2018)\citenamefont
  {Cornelissen}, \citenamefont {Liu}, \citenamefont {van Wees},\ and\
  \citenamefont {Duine}}]{Cornelissen_2018}%
  \BibitemOpen
  \bibfield  {author} {\bibinfo {author} {\bibfnamefont {L.}~\bibnamefont
  {Cornelissen}}, \bibinfo {author} {\bibfnamefont {J.}~\bibnamefont {Liu}},
  \bibinfo {author} {\bibfnamefont {B.}~\bibnamefont {van Wees}}, \ and\
  \bibinfo {author} {\bibfnamefont {R.}~\bibnamefont {Duine}},\ }\bibfield
  {title} {\enquote {\bibinfo {title} {Spin-current-controlled modulation of
  the magnon spin conductance in a three-terminal magnon transistor},}\ }\href
  {\doibase 10.1103/physrevlett.120.097702} {\bibfield  {journal} {\bibinfo
  {journal} {Physical Review Letters}\ }\textbf {\bibinfo {volume} {120}}
  (\bibinfo {year} {2018}),\ 10.1103/physrevlett.120.097702}\BibitemShut
  {NoStop}%
\bibitem [{\citenamefont {Morris}\ \emph {et~al.}(2017)\citenamefont {Morris},
  \citenamefont {van Loo}, \citenamefont {Kosen},\ and\ \citenamefont
  {Karenowska}}]{Morris_2017}%
  \BibitemOpen
  \bibfield  {author} {\bibinfo {author} {\bibfnamefont {R.~G.~E.}\
  \bibnamefont {Morris}}, \bibinfo {author} {\bibfnamefont {A.~F.}\
  \bibnamefont {van Loo}}, \bibinfo {author} {\bibfnamefont {S.}~\bibnamefont
  {Kosen}}, \ and\ \bibinfo {author} {\bibfnamefont {A.~D.}\ \bibnamefont
  {Karenowska}},\ }\bibfield  {title} {\enquote {\bibinfo {title} {{Strong
  coupling of magnons in a YIG sphere to photons in a planar superconducting
  resonator in the quantum limit}},}\ }\href {\doibase
  10.1038/s41598-017-11835-4} {\bibfield  {journal} {\bibinfo  {journal}
  {Scientific Reports}\ }\textbf {\bibinfo {volume} {7}} (\bibinfo {year}
  {2017}),\ 10.1038/s41598-017-11835-4}\BibitemShut {NoStop}%
\bibitem [{\citenamefont {Zhang}\ \emph {et~al.}(2014)\citenamefont {Zhang},
  \citenamefont {Zou}, \citenamefont {Jiang},\ and\ \citenamefont
  {Tang}}]{Zhang_2014}%
  \BibitemOpen
  \bibfield  {author} {\bibinfo {author} {\bibfnamefont {X.}~\bibnamefont
  {Zhang}}, \bibinfo {author} {\bibfnamefont {C.-L.}\ \bibnamefont {Zou}},
  \bibinfo {author} {\bibfnamefont {L.}~\bibnamefont {Jiang}}, \ and\ \bibinfo
  {author} {\bibfnamefont {H.~X.}\ \bibnamefont {Tang}},\ }\bibfield  {title}
  {\enquote {\bibinfo {title} {Strongly coupled magnons and cavity microwave
  photons},}\ }\href {\doibase 10.1103/physrevlett.113.156401} {\bibfield
  {journal} {\bibinfo  {journal} {Physical Review Letters}\ }\textbf {\bibinfo
  {volume} {113}} (\bibinfo {year} {2014}),\
  10.1103/physrevlett.113.156401}\BibitemShut {NoStop}%
\bibitem [{\citenamefont {LeCraw}, \citenamefont {Spencer},\ and\ \citenamefont
  {Porter}(1958)}]{LeCraw_1958}%
  \BibitemOpen
  \bibfield  {author} {\bibinfo {author} {\bibfnamefont {R.~C.}\ \bibnamefont
  {LeCraw}}, \bibinfo {author} {\bibfnamefont {E.~G.}\ \bibnamefont {Spencer}},
  \ and\ \bibinfo {author} {\bibfnamefont {C.~S.}\ \bibnamefont {Porter}},\
  }\bibfield  {title} {\enquote {\bibinfo {title} {Ferromagnetic resonance line
  width in yttrium iron garnet single crystals},}\ }\href {\doibase
  10.1103/physrev.110.1311} {\bibfield  {journal} {\bibinfo  {journal}
  {Physical Review}\ }\textbf {\bibinfo {volume} {110}},\ \bibinfo {pages}
  {1311–1313} (\bibinfo {year} {1958})}\BibitemShut {NoStop}%
\end{thebibliography}%


\begin{thebibliography}{11}%
\makeatletter
\providecommand \@ifxundefined [1]{%
 \@ifx{#1\undefined}
}%
\providecommand \@ifnum [1]{%
 \ifnum #1\expandafter \@firstoftwo
 \else \expandafter \@secondoftwo
 \fi
}%
\providecommand \@ifx [1]{%
 \ifx #1\expandafter \@firstoftwo
 \else \expandafter \@secondoftwo
 \fi
}%
\providecommand \natexlab [1]{#1}%
\providecommand \enquote  [1]{``#1''}%
\providecommand \bibnamefont  [1]{#1}%
\providecommand \bibfnamefont [1]{#1}%
\providecommand \citenamefont [1]{#1}%
\providecommand \href@noop [0]{\@secondoftwo}%
\providecommand \href [0]{\begingroup \@sanitize@url \@href}%
\providecommand \@href[1]{\@@startlink{#1}\@@href}%
\providecommand \@@href[1]{\endgroup#1\@@endlink}%
\providecommand \@sanitize@url [0]{\catcode `\\12\catcode `\$12\catcode
  `\&12\catcode `\#12\catcode `\^12\catcode `\_12\catcode `\%12\relax}%
\providecommand \@@startlink[1]{}%
\providecommand \@@endlink[0]{}%
\providecommand \url  [0]{\begingroup\@sanitize@url \@url }%
\providecommand \@url [1]{\endgroup\@href {#1}{\urlprefix }}%
\providecommand \urlprefix  [0]{URL }%
\providecommand \Eprint [0]{\href }%
\providecommand \doibase [0]{http://dx.doi.org/}%
\providecommand \selectlanguage [0]{\@gobble}%
\providecommand \bibinfo  [0]{\@secondoftwo}%
\providecommand \bibfield  [0]{\@secondoftwo}%
\providecommand \translation [1]{[#1]}%
\providecommand \BibitemOpen [0]{}%
\providecommand \bibitemStop [0]{}%
\providecommand \bibitemNoStop [0]{.\EOS\space}%
\providecommand \EOS [0]{\spacefactor3000\relax}%
\providecommand \BibitemShut  [1]{\csname bibitem#1\endcsname}%
\let\auto@bib@innerbib\@empty
\bibitem [{\citenamefont {Yu}\ \emph {et~al.}(2014)\citenamefont {Yu},
  \citenamefont {Harberts}, \citenamefont {Adur}, \citenamefont {Lu},
  \citenamefont {Hammel}, \citenamefont {Johnston-Halperin},\ and\
  \citenamefont {Epstein}}]{Yu_2014}%
  \BibitemOpen
  \bibfield  {author} {\bibinfo {author} {\bibfnamefont {H.}~\bibnamefont
  {Yu}}, \bibinfo {author} {\bibfnamefont {M.}~\bibnamefont {Harberts}},
  \bibinfo {author} {\bibfnamefont {R.}~\bibnamefont {Adur}}, \bibinfo {author}
  {\bibfnamefont {Y.}~\bibnamefont {Lu}}, \bibinfo {author} {\bibfnamefont
  {P.~C.}\ \bibnamefont {Hammel}}, \bibinfo {author} {\bibfnamefont
  {E.}~\bibnamefont {Johnston-Halperin}}, \ and\ \bibinfo {author}
  {\bibfnamefont {A.~J.}\ \bibnamefont {Epstein}},\ }\bibfield  {title}
  {\enquote {\bibinfo {title} {Ultra-narrow ferromagnetic resonance in
  organic-based thin films grown via low temperature chemical vapor
  deposition},}\ }\href {\doibase 10.1063/1.4887924} {\bibfield  {journal}
  {\bibinfo  {journal} {Applied Physics Letters}\ }\textbf {\bibinfo {volume}
  {105}},\ \bibinfo {pages} {012407} (\bibinfo {year} {2014})}\BibitemShut
  {NoStop}%
\bibitem [{\citenamefont {Vansteenkiste}\ \emph {et~al.}(2014)\citenamefont
  {Vansteenkiste}, \citenamefont {Leliaert}, \citenamefont {Dvornik},
  \citenamefont {Helsen}, \citenamefont {Garcia-Sanchez},\ and\ \citenamefont
  {Van~Waeyenberge}}]{Vansteenkiste_2014}%
  \BibitemOpen
  \bibfield  {author} {\bibinfo {author} {\bibfnamefont {A.}~\bibnamefont
  {Vansteenkiste}}, \bibinfo {author} {\bibfnamefont {J.}~\bibnamefont
  {Leliaert}}, \bibinfo {author} {\bibfnamefont {M.}~\bibnamefont {Dvornik}},
  \bibinfo {author} {\bibfnamefont {M.}~\bibnamefont {Helsen}}, \bibinfo
  {author} {\bibfnamefont {F.}~\bibnamefont {Garcia-Sanchez}}, \ and\ \bibinfo
  {author} {\bibfnamefont {B.}~\bibnamefont {Van~Waeyenberge}},\ }\bibfield
  {title} {\enquote {\bibinfo {title} {{The design and verification of
  MuMax3}},}\ }\href {\doibase 10.1063/1.4899186} {\bibfield  {journal}
  {\bibinfo  {journal} {AIP Advances}\ }\textbf {\bibinfo {volume} {4}},\
  \bibinfo {pages} {107133} (\bibinfo {year} {2014})}\BibitemShut {NoStop}%
\bibitem [{\citenamefont {Kalinikos}\ and\ \citenamefont
  {Slavin}(1986)}]{Kalinikos_1986}%
  \BibitemOpen
  \bibfield  {author} {\bibinfo {author} {\bibfnamefont {B.~A.}\ \bibnamefont
  {Kalinikos}}\ and\ \bibinfo {author} {\bibfnamefont {A.~N.}\ \bibnamefont
  {Slavin}},\ }\bibfield  {title} {\enquote {\bibinfo {title} {Theory of
  dipole-exchange spin wave spectrum for ferromagnetic films with mixed
  exchange boundary conditions},}\ }\href {\doibase
  10.1088/0022-3719/19/35/014} {\bibfield  {journal} {\bibinfo  {journal}
  {Journal of Physics C: Solid State Physics}\ }\textbf {\bibinfo {volume}
  {19}},\ \bibinfo {pages} {7013–7033} (\bibinfo {year} {1986})}\BibitemShut
  {NoStop}%
\bibitem [{\citenamefont {Foreman-Mackey}\ \emph {et~al.}(2013)\citenamefont
  {Foreman-Mackey}, \citenamefont {Hogg}, \citenamefont {Lang},\ and\
  \citenamefont {Goodman}}]{Foreman_Mackey_2013}%
  \BibitemOpen
  \bibfield  {author} {\bibinfo {author} {\bibfnamefont {D.}~\bibnamefont
  {Foreman-Mackey}}, \bibinfo {author} {\bibfnamefont {D.~W.}\ \bibnamefont
  {Hogg}}, \bibinfo {author} {\bibfnamefont {D.}~\bibnamefont {Lang}}, \ and\
  \bibinfo {author} {\bibfnamefont {J.}~\bibnamefont {Goodman}},\ }\bibfield
  {title} {\enquote {\bibinfo {title} {{emcee: The MCMC hammer}},}\ }\href
  {\doibase 10.1086/670067} {\bibfield  {journal} {\bibinfo  {journal}
  {Publications of the Astronomical Society of the Pacific}\ }\textbf {\bibinfo
  {volume} {125}},\ \bibinfo {pages} {306–312} (\bibinfo {year}
  {2013})}\BibitemShut {NoStop}%
\bibitem [{\citenamefont {Newville}\ \emph {et~al.}(2014)\citenamefont
  {Newville}, \citenamefont {Stensitzki}, \citenamefont {Allen},\ and\
  \citenamefont {Ingargiola}}]{Newville_Matthew_2014_11813}%
  \BibitemOpen
  \bibfield  {author} {\bibinfo {author} {\bibfnamefont {M.}~\bibnamefont
  {Newville}}, \bibinfo {author} {\bibfnamefont {T.}~\bibnamefont
  {Stensitzki}}, \bibinfo {author} {\bibfnamefont {D.~B.}\ \bibnamefont
  {Allen}}, \ and\ \bibinfo {author} {\bibfnamefont {A.}~\bibnamefont
  {Ingargiola}},\ }\href {\doibase 10.5281/zenodo.11813} {\enquote {\bibinfo
  {title} {{LMFIT: Non-linear least-square minimization and curve-fitting for
  Python}},}\ } (\bibinfo {year} {2014})\BibitemShut {NoStop}%
\bibitem [{\citenamefont {Kalarickal}\ \emph {et~al.}(2006)\citenamefont
  {Kalarickal}, \citenamefont {Krivosik}, \citenamefont {Wu}, \citenamefont
  {Patton}, \citenamefont {Schneider}, \citenamefont {Kabos}, \citenamefont
  {Silva},\ and\ \citenamefont {Nibarger}}]{Kalarickal_2006}%
  \BibitemOpen
  \bibfield  {author} {\bibinfo {author} {\bibfnamefont {S.~S.}\ \bibnamefont
  {Kalarickal}}, \bibinfo {author} {\bibfnamefont {P.}~\bibnamefont
  {Krivosik}}, \bibinfo {author} {\bibfnamefont {M.}~\bibnamefont {Wu}},
  \bibinfo {author} {\bibfnamefont {C.~E.}\ \bibnamefont {Patton}}, \bibinfo
  {author} {\bibfnamefont {M.~L.}\ \bibnamefont {Schneider}}, \bibinfo {author}
  {\bibfnamefont {P.}~\bibnamefont {Kabos}}, \bibinfo {author} {\bibfnamefont
  {T.~J.}\ \bibnamefont {Silva}}, \ and\ \bibinfo {author} {\bibfnamefont
  {J.~P.}\ \bibnamefont {Nibarger}},\ }\bibfield  {title} {\enquote {\bibinfo
  {title} {{Ferromagnetic resonance linewidth in metallic thin films:
  Comparison of measurement methods}},}\ }\href {\doibase 10.1063/1.2197087}
  {\bibfield  {journal} {\bibinfo  {journal} {Journal of Applied Physics}\
  }\textbf {\bibinfo {volume} {99}},\ \bibinfo {pages} {093909} (\bibinfo
  {year} {2006})}\BibitemShut {NoStop}%
\bibitem [{\citenamefont {Sparks}(1970)}]{Sparks_1970}%
  \BibitemOpen
  \bibfield  {author} {\bibinfo {author} {\bibfnamefont {M.}~\bibnamefont
  {Sparks}},\ }\bibfield  {title} {\enquote {\bibinfo {title} {Magnetostatic
  modes in an infinite circular disk},}\ }\href {\doibase
  10.1016/0038-1098(70)90419-9} {\bibfield  {journal} {\bibinfo  {journal}
  {Solid State Communications}\ }\textbf {\bibinfo {volume} {8}},\ \bibinfo
  {pages} {731–733} (\bibinfo {year} {1970})}\BibitemShut {NoStop}%
\bibitem [{\citenamefont {Candido}\ and\ \citenamefont
  {Flatt\'e}()}]{denisflatte}%
  \BibitemOpen
  \bibfield  {author} {\bibinfo {author} {\bibfnamefont {D.~R.}\ \bibnamefont
  {Candido}}\ and\ \bibinfo {author} {\bibfnamefont {M.~E.}\ \bibnamefont
  {Flatt\'e}},\ }\href@noop {} {\bibinfo  {journal} {to be published}\
  }\BibitemShut {NoStop}%
\bibitem [{\citenamefont {Kakazei}\ \emph {et~al.}(2004)\citenamefont
  {Kakazei}, \citenamefont {Wigen}, \citenamefont {Guslienko}, \citenamefont
  {Novosad}, \citenamefont {Slavin}, \citenamefont {Golub}, \citenamefont
  {Lesnik},\ and\ \citenamefont {Otani}}]{Kakazei_2004}%
  \BibitemOpen
\bibfield  {journal} {  }\bibfield  {author} {\bibinfo {author} {\bibfnamefont
  {G.~N.}\ \bibnamefont {Kakazei}}, \bibinfo {author} {\bibfnamefont {P.~E.}\
  \bibnamefont {Wigen}}, \bibinfo {author} {\bibfnamefont {K.~Y.}\ \bibnamefont
  {Guslienko}}, \bibinfo {author} {\bibfnamefont {V.}~\bibnamefont {Novosad}},
  \bibinfo {author} {\bibfnamefont {A.~N.}\ \bibnamefont {Slavin}}, \bibinfo
  {author} {\bibfnamefont {V.~O.}\ \bibnamefont {Golub}}, \bibinfo {author}
  {\bibfnamefont {N.~A.}\ \bibnamefont {Lesnik}}, \ and\ \bibinfo {author}
  {\bibfnamefont {Y.}~\bibnamefont {Otani}},\ }\href {\doibase
  10.1063/1.1772868} {\enquote {\bibinfo {title} {Spin-wave spectra of
  perpendicularly magnetized circular submicron dot arrays},}\ } (\bibinfo
  {year} {2004})\BibitemShut {NoStop}%
\bibitem [{\citenamefont {Nedukh}\ \emph {et~al.}(2013)\citenamefont {Nedukh},
  \citenamefont {Tarapov}, \citenamefont {Belozorov}, \citenamefont
  {Kharchenko}, \citenamefont {Golub}, \citenamefont {Kilimchuk}, \citenamefont
  {Salyuk}, \citenamefont {Tartakovskaya}, \citenamefont {Bunyaev},\ and\
  \citenamefont {Kakazei}}]{Nedukh_2013}%
  \BibitemOpen
  \bibfield  {author} {\bibinfo {author} {\bibfnamefont {S.~V.}\ \bibnamefont
  {Nedukh}}, \bibinfo {author} {\bibfnamefont {S.~I.}\ \bibnamefont {Tarapov}},
  \bibinfo {author} {\bibfnamefont {D.~P.}\ \bibnamefont {Belozorov}}, \bibinfo
  {author} {\bibfnamefont {A.~A.}\ \bibnamefont {Kharchenko}}, \bibinfo
  {author} {\bibfnamefont {V.~O.}\ \bibnamefont {Golub}}, \bibinfo {author}
  {\bibfnamefont {I.~V.}\ \bibnamefont {Kilimchuk}}, \bibinfo {author}
  {\bibfnamefont {O.~Y.}\ \bibnamefont {Salyuk}}, \bibinfo {author}
  {\bibfnamefont {E.~V.}\ \bibnamefont {Tartakovskaya}}, \bibinfo {author}
  {\bibfnamefont {S.~A.}\ \bibnamefont {Bunyaev}}, \ and\ \bibinfo {author}
  {\bibfnamefont {G.~N.}\ \bibnamefont {Kakazei}},\ }\bibfield  {title}
  {\enquote {\bibinfo {title} {Standing spin waves in perpendicularly
  magnetized circular dots at millimeter waves},}\ }\href {\doibase
  10.1063/1.4799528} {\bibfield  {journal} {\bibinfo  {journal} {Journal of
  Applied Physics}\ }\textbf {\bibinfo {volume} {113}},\ \bibinfo {pages}
  {17B521} (\bibinfo {year} {2013})}\BibitemShut {NoStop}%
\bibitem [{\citenamefont {Maksymov}\ and\ \citenamefont
  {Kostylev}(2015)}]{Maksymov_2015}%
  \BibitemOpen
  \bibfield  {author} {\bibinfo {author} {\bibfnamefont {I.~S.}\ \bibnamefont
  {Maksymov}}\ and\ \bibinfo {author} {\bibfnamefont {M.}~\bibnamefont
  {Kostylev}},\ }\bibfield  {title} {\enquote {\bibinfo {title} {Broadband
  stripline ferromagnetic resonance spectroscopy of ferromagnetic films,
  multilayers and nanostructures},}\ }\href {\doibase
  10.1016/j.physe.2014.12.027} {\bibfield  {journal} {\bibinfo  {journal}
  {Physica E: Low-dimensional Systems and Nanostructures}\ }\textbf {\bibinfo
  {volume} {69}},\ \bibinfo {pages} {253–293} (\bibinfo {year}
  {2015})}\BibitemShut {NoStop}%
\end{thebibliography}%

\end{document}


\title{Low-Damping Ferromagnetic Resonance in Electron-Beam Patterned, High-\(Q\) Vanadium Tetracyanoethylene Magnon Cavities}

\author{Andrew Franson}
\affiliation{Department of Physics, The Ohio State University, Columbus, Ohio 43210, USA}
\author{Na Zhu}
\affiliation{Department of Electrical Engineering, Yale University, New Haven, Connecticut 06511, USA}
\author{Seth Kurfman}
\author{Michael Chilcote}
\affiliation{Department of Physics, The Ohio State University, Columbus, Ohio 43210, USA}
\author{Denis R. Candido}
\affiliation{Department of Physics and Astronomy, University of Iowa, Iowa City, Iowa 52242, USA}
\affiliation{Pritzker School of Molecular Engineering, University of Chicago, Chicago, Illinois 60637, USA}
\author{Kristen S. Buchanan}
\affiliation{Department of Physics, Colorado State University, Fort Collins, Colorado 80523, USA}
\author{Michael E. Flatt\'e}
\affiliation{Department of Physics and Astronomy, University of Iowa, Iowa City, Iowa 52242, USA}
\affiliation{Pritzker School of Molecular Engineering, University of Chicago, Chicago, Illinois 60637, USA}
\author{Hong X. Tang}
\affiliation{Department of Electrical Engineering, Yale University, New Haven, Connecticut 06511, USA}
\author{Ezekiel Johnston-Halperin}
\affiliation{Department of Physics, The Ohio State University, Columbus, Ohio 43210, USA}
\email{johnston.halperin@gmail.com}

\maketitle

\section*{Patterning and Measurement Methods}
\subsection*{Patterning and Growth}
\VTCNE bars and disks are  patterned on commercially available C-plane
polished  sapphire  (\symb{Al_2O_3})  via  electron-beam  lithographic
techniques.  The  substrates  are  cleaned with  a  solvent  chain  of
acetone, methanol,  isopropanol (IPA), and deionized  water (DI water)
followed by  a 20  minute ultraviolet  ozone clean  (UVOC) in  a UVOCS
T10x10/OES  to degrease  and remove  organic contaminants.  A 400  nm
layer of MMA (8.5)  MAA EL 11 (P(MMA-MAA)) is spun on  at 2000 rpm for
45 seconds  then soft baked  at \SI{180}{\celsius} for 300  seconds. A
140 nm layer of  495PMMA A6 (PMMA) is then spun on at  2000 rpm for 45
seconds then soft baked at \SI{180}{\celsius}  for 60 seconds. A 10 nm
thick  layer  of  aluminum  is deposited  via  thermal  deposition  at
\num{1e-6} Torr.  The electron-beam patterning of  the PMMA/P(MMA-MAA)
bilayer is performed on a FEI Helios Nanolab 600 Dual Beam Focused Ion
Beam/Scanning  Electron Microscope  with the  assistance of  Nanometer
Pattern Generation System (NPGS)  software. Development of the written
pattern is  achieved with Microposit  MF-319 for 40 seconds,  DI water
for  20 seconds,  MF-319  for 40  seconds, DI  water  for 20  seconds,
Microchem MIBK:IPA (1:3) for 60 seconds,  IPA for 20 seconds, DI water
for  20  seconds. That  is  followed  by a  120  second  hard bake  at
\SI{100}{\celsius}. A 3  nm thick layer of aluminum  is then deposited
in the same system  as the prior 10 nm layer  to prevent outgassing of
the PMMA/P(MMA-MAA)  bilayer during  \VTCNE deposition. The  sample is
oxidized and cleaned with a  10-minute UVOC. This oxidizes the surface
of the  3 nm  aluminum layer  and removes  potential small  sources of
contamination from the growth surfaces.  Growth of \VTCNE is described
in previous work.\cite{Yu_2014}  After transfer of the  films from the
growth  glovebox to  the  liftoff glovebox,  liftoff  is performed  in
dichloromethane.  For the  feature  sizes and  thicknesses used  here,
liftoff occurred in a few minutes with gentle agitation from a Pasteur
pipette.

\subsection*{Microwave Measurements}
Angle-resolved  ferromagnetic resonance  (FMR)  measurements are  done
with  a Bruker  electron paramagnetic  resonance spectrometer  with an
X-band bridge and 10 kOe electromagnet. Before measurement, the \VTCNE
samples  are sealed  into a  quartz tube  with a  ceramic holder  that
aligns  the  normal plane  of  the  sample.   Between each  scan,  the
microwave  frequency is  tuned  between  9 and  10  GHz  to match  the
resonant frequency  of the loaded  cavity. The frequency is  fixed and
measurements  are then  performed by  sweeping the  static field  with
\SI{200}{\micro  \watt} of  applied microwave  power and  a modulation
field of 0.1 Oe.  The quartz tube  has a pointer fixed to it, allowing
for  alignment within  0.5 degrees  of the  sample with  respect to  a
custom made goniometer.  The larger error of \(\pm 2\) degrees seen in
Fig.  3(d)  comes from  the initial aligning  the goniometer  with the
in-plane (IP, \(\theta=90\))  orientation of the \VTCNE  film.  The IP
orientation is taken  to be the point where the  resonance field is at
its minimum.

Frequency-resolved microwave  measurements are  done with  a broadband
ferromagnetic  resonance (BFMR)  setup with  a B4003-8M-50  microstrip
test  board from  Southwest Microwave  that is  sourced by  an Agilent
N5222A vector  network analyzer (VNA),  transduced via a  Krytar 203BK
Schottky  diode, and  measured  with  an Ametek  7265  Dual Phase  DSP
lock-in  amplifier.  The  microstrip  is positioned  inside  a 10  kOe
electromagnet   in   the   out-of-plane  (OOP,   \(\theta=0\))   field
geometry. Measurements  are performed with  an input power of  -10 dBm
and a modulation field of roughly 0.1 Oe oscillating at 577 Hz.

\newpage
\section*{Numerical Modeling}
Micromagnetic simulations are  done to gain insight  into the measured
FMR   spectra.   The   micromagnetic   simulations   are  done   using
MuMax3\cite{Vansteenkiste_2014} with the  following parameters for the
\VTCNEs: saturation magnetization \symb{4 \pi M_s} \(=\) \SI{76.57}{G}
(\(6093\)     A/m),     exchange     constant     \(A_{ex}\)     \(=\)
\SI{2.2e-10}{erg\per\cm}   (\SI{2.2e-15}{\J/\m}),    and   a   damping
parameter of  \(\alpha=0.0001\), where  \(\alpha\) is larger  than the
smallest measured  damping value for  \VTCNE and was chosen  to ensure
that the  simulations would converge  in a reasonable amount  of time.
Cells  of  40  x  40   x  4.6875  \symb{nm^{3}}  are  used.   Selected
simulations  are  repeated with  smaller  cells  with similar  results
because  the  40  nm  cell  sizes are  still  small  compared  to  the
wavelengths  of  the spin-wave  modes.   \(A_{ex}\)  is determined  by
fitting the mode spacing of the  thickness confined modes from the OOP
spectra of the disks blue anisotropy curves in Fig.  3(c).  Fitting is
done to theory from Kalinikos et al.\cite{Kalinikos_1986} and is shown
in Fig.  3(e). Fitting for  fully pinned and fully unpinned geometries
yielded the same value for \(A_{ex}\).

The simulations are done with  the static field, \(H_{bias}\), applied
OOP.  The  field  magnitudes  are  chosen to  result  in  a  resonance
frequency near  9.83 GHz  to be similar  to the  measurement frequency
used in the experiments. The structures are relaxed in the presence of
\(H_{bias}\) and a small additional  field \(H_{dynam}\) of roughly 50
Oe,  chosen such  that the  spins tilt  by about  1\% from  the static
equilibrium position, is applied in the x-direction.  The dynamics are
monitored as a  function of time after removing  \(H_{dynam}\) and the
simulated spectra are obtained by taking the Fourier transforms of the
x-component  of the  magnetization  for each  run.  Mode profiles  are
calculated  for  selected  peaks  in the  spectra  by  running  driven
simulations  at the  selected resonance  frequency and  extracting the
spin distributions as  a function of time for a  full period after the
simulation has reached a steady state.

\begin{figure}
    \includegraphics[width=0.31\textwidth]{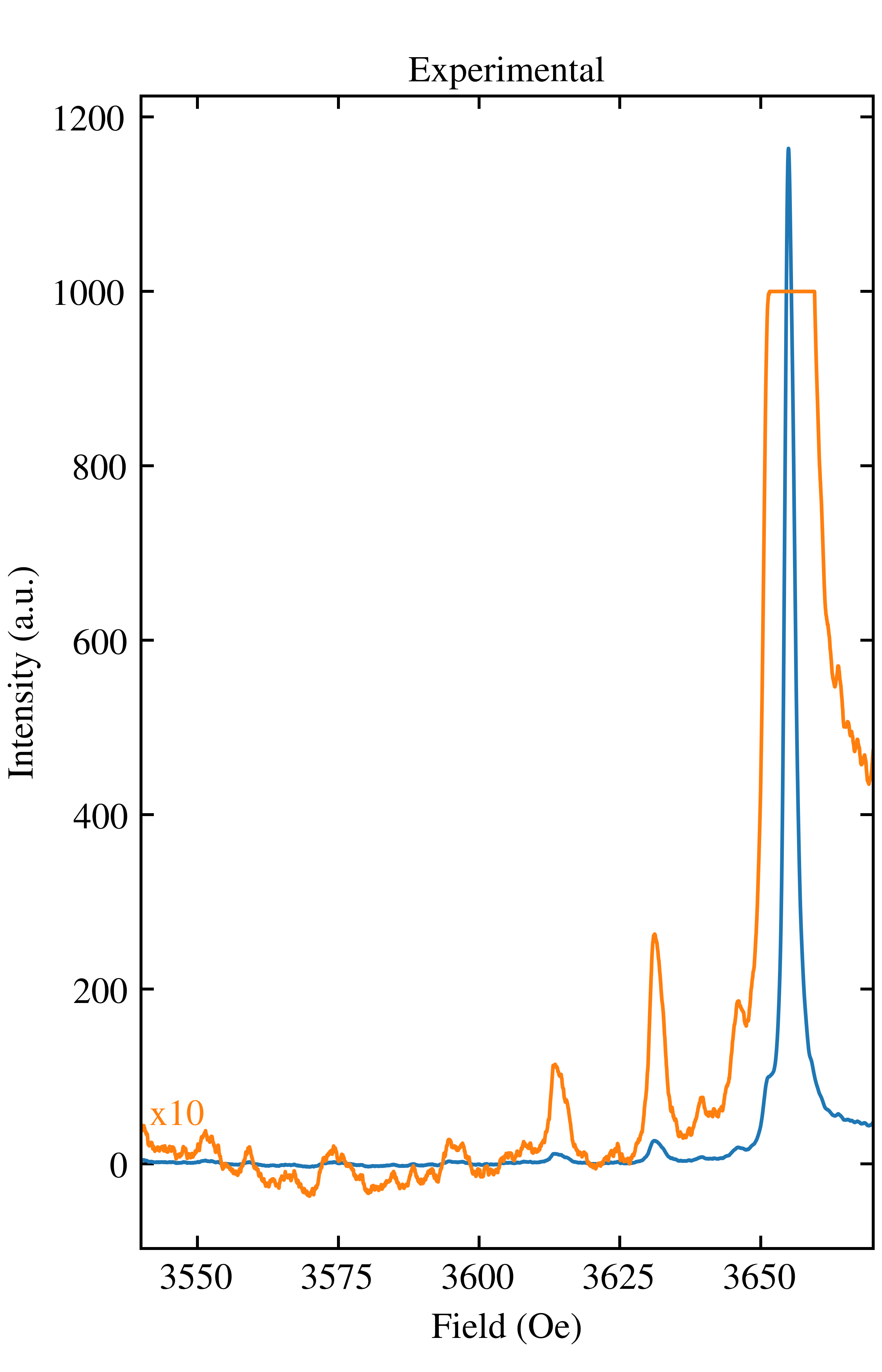}
    \includegraphics[width=0.62\textwidth]{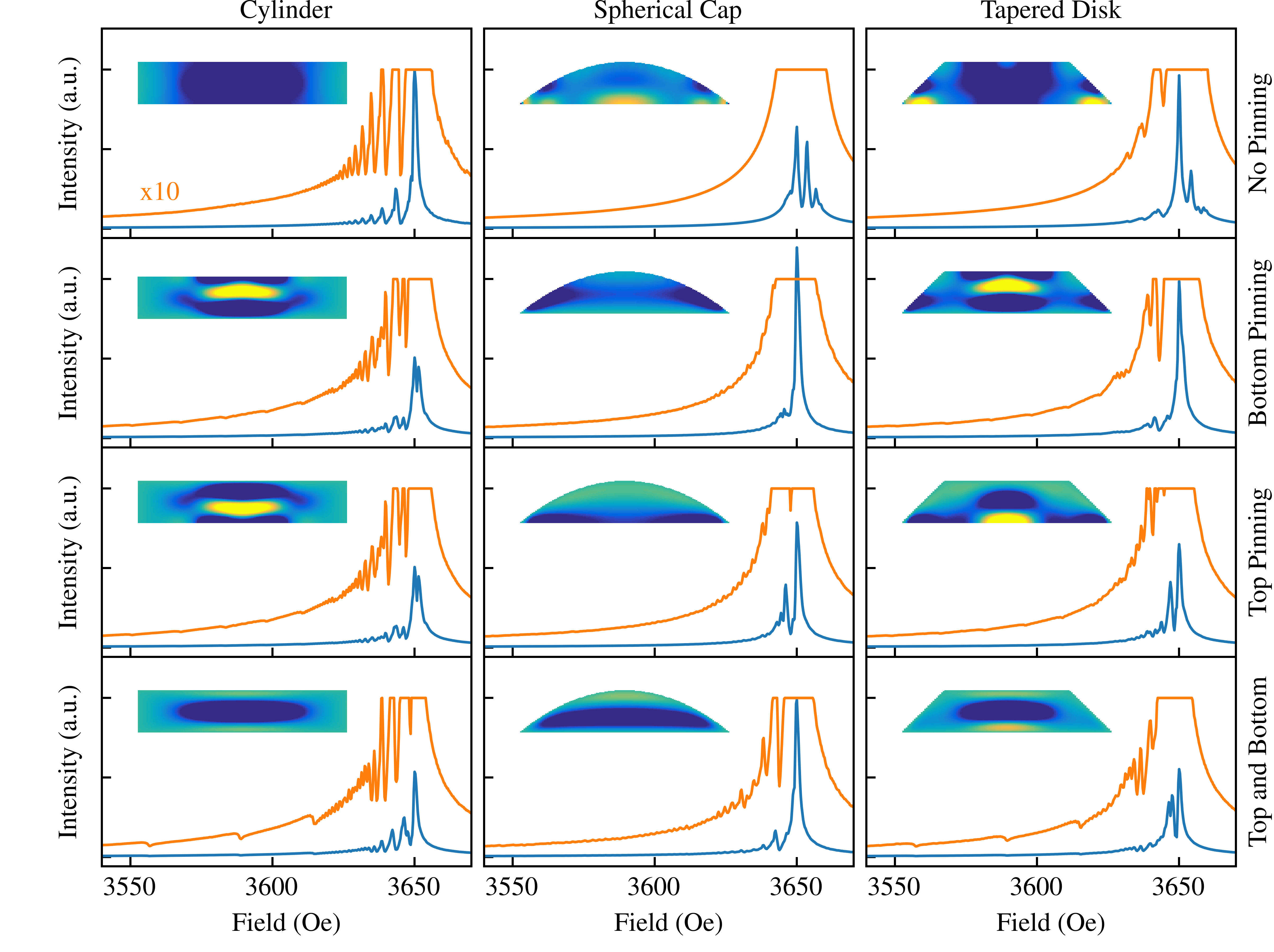}
    \caption{Left, experimental data  of the 9.83 GHz OOP  FMR scan of
        the array of \SI{5}{\um} disks.  Right, array of the different
        simulated  geometries considered  in this  study.  Blue  lines
        represent the full data and the orange lines are the full data
        multiplied by  10 and truncated to  better show low-intensity,
        low-field behavior.  Mode maps  of the vertical cross-sections
        of the mode maps at the  disk center for the largest amplitude
        peak for each  simulated geometry are shown in  the upper left
        of  each plot,  where dark  blue (dark),  green (medium),  and
        yellow (light) represent negative,  zero, and positive motion,
        respectively,  at an  instant  of time.   These are  quantized
        standing modes.}
    \label{fig:comparisons}
\end{figure}

Since the pinning  of the structures is unknown and  the exact profile
is  uncertain,  a matrix  of  different  combinations of  pinning  and
cross-sectional  profile   is  investigated.   Three   geometries  are
considered  based on  the growth  characteristics of  the bar  in Fig.
1(b,  c), a  cylindrical disk;  a  lens-shaped, circular  disk with  a
spherically-curved  top surface  and  flat  bottom surface  (spherical
cap); and a  disk with a \SI{1}{\um}  wide ramp from the  outer rim to
the  inner   rim,  Fig.~\ref{fig:comparisons}.   For  each   of  these
geometries,  simulations with  no  pinning of  surface spins,  perfect
pinning on the top surface, perfect pinning on the bottom surface, and
perfect pinning on both the top and bottom surfaces is considered.  To
compare the  experimental spectra with the  simulations, we considered
the prevalence of  the thickness modes and the shape  of the strongest
peak. The mode spacing does not change appreciably with the disk shape
(taper vs.  lens),  and agrees well with the calculations  of the mode
spacing for an  unpatterned thin film. The spherical cap  used for the
lens-type  simulations leads  to  almost complete  suppression of  the
thickness  modes for  all  of  the pinning  conditions,  which is  not
consistent with the  experimental spectrum. The strongest  peak in the
data is smoother than the jagged combo-like structure of the main peak
of the  cylindrical simulations that  occurs due to the  radial modes.
The two simulations that are the  closest to the experimental data are
the   tapered-disk  simulations   with   top/bottom  and   bottom-only
pinning. The  even modes show up  almost as strongly as  the odd modes
for bottom-only  pinning, whereas only  the odd modes are  present for
top/bottom  pinning.  As  show  in Fig.   3(e),  resonance peaks  that
correspond to  even thickness modes  are present but weak  compared to
the odd modes,  which suggests that both surfaces are  pinned but that
pinning is imperfect on one of the two surfaces, likely the top.

\newpage
\section*{Fitting Methods}
All    fitting   is    performed    in   Python    with   the    emcee
package\cite{Foreman_Mackey_2013}        within       the        lmfit
package.\cite{Newville_Matthew_2014_11813}

\subsection*{Cavity FMR Angular Anisotropy Fitting}
Fitting of  the red anisotropy curves  in Fig. 3(d) are  all performed
with a  modified Eq 2. Squaring  both sides and collecting  factors of
\(H\) yields
\begin{multline}
    \left(\frac{\omega}{\gamma}\right)^2 =
    H^2 + 
    H \times 4 \pi M_s\left(- N_{IP} \cos(2\theta) - N_{IP} \cos(\theta)^2 - N_{OP} \cos(2\varphi)\right) + \\
    16 \pi^2 M_s^2 \left( N_{IP}^2 \cos(\theta)^2 + N_{IP} N_{OP} \cos(2\theta) \cos(2 \varphi) -
    N_{OP}^2 \cos(\theta)^2 \cos(\varphi)^2 \sin(\varphi)^2\right),
\end{multline}
solving for \(H\) then yields
\begin{equation}
    H = \frac{-b \pm \sqrt{b^2 - 4 a c}}{2 a}
\end{equation}
where
\begin{subequations}
    \begin{equation}
        a \equiv 1,
    \end{equation}
    \begin{equation}
        b \equiv 4 \pi M_s(- N_{IP} \cos(2\theta) - N_{IP} \cos(\theta)^2 - N_{OP} \cos(2\varphi)),
    \end{equation}
    \begin{multline}
        c \equiv 16 \pi^2 M_s^2 \left( N_{IP}^2 \cos(\theta)^2 + N_{IP} N_{OP} \cos(2\theta) \cos(2 \varphi) - \right.\\
    \left. N_{OP}^2 \cos(\theta)^2 \cos(\varphi)^2 \sin(\varphi)^2\right) - \left(\frac{\omega}{\gamma}\right)^2.
    \end{multline}
\end{subequations}
To account  for deviations  from high-symmetry  directions, \(\theta\)
and  \(\varphi\)  are  parameterized  in terms  of  a  new  parameter,
\(t\). This parameterization  allows for fitting through  a path along
any great circle of the unit sphere by using the expression
\begin{subequations}
    \begin{equation}
        \vec{v_1} = \sin(\theta_1) \cos(\varphi_1) \hat{x} + \sin(\theta_1) \sin(\varphi_1) \hat{y} + \cos(\theta_1) \hat{z}
    \end{equation}
    \begin{equation}
        \vec{v_2} = \sin(\theta_2) \cos(\varphi_2) \hat{x} + \sin(\theta_2) \sin(\varphi_2) \hat{y} + \cos(\theta_2) \hat{z}
    \end{equation}
    \begin{equation}
        \eta = \arccos(\vec{v_1} \cdot \vec{v_2})
    \end{equation}
    \begin{equation}
        \vec{v_m} = \frac{\sin(\eta (1-t)) \vec{v_1} + \sin(\eta t) \vec{v_2}}{\sin(\eta)}
    \end{equation}
\end{subequations}
where  \(\vec{v_m}\) points  to a  location on  the great  circle that
intersects \(\vec{v_1}\) and \(\vec{v_2}\). The location is determined
by  the   parameter  \(t\)   which  steps  from   \(t_{i}  =   0\)  to
\(t_{f}       =       \frac{\pi}{\eta}\)       in       steps       of
\(t_{step} =  \frac{t_{f} -  t_{i}}{18}\) to  produce \SI{10}{\degree}
steps from \SI{0}{\degree} to \SI{180}{\degree}.

This correction is only necessary for the bar sample scanned IP to OOP
with the applied field perpendicular to  the bar axis (red diamonds in
Fig.  3(d)).   A path  traveling from  the IP  position (\(\theta_1=\)
\SI{90}{\degree},  \(\varphi_1=\)  \SI{90}{\degree})   to  a  position
\SI{10}{\degree}  away   from  OOP   (\(\theta_2=\)  \SI{10}{\degree},
\(\varphi_2=\)  \SI{90}{\degree}) is  required to  accurately describe
the data.   This corresponds  to an initial  \(\theta\) offset  in the
x-direction while sweeping \(\theta\)  in the y-direction and explains
why the red-diamond curve does  not kiss the red-circle and red-square
curves at exactly one point.

\subsection*{Broadband FMR Linewidth Fitting}

\begin{figure}[!htbp]
  \centering
  \includegraphics{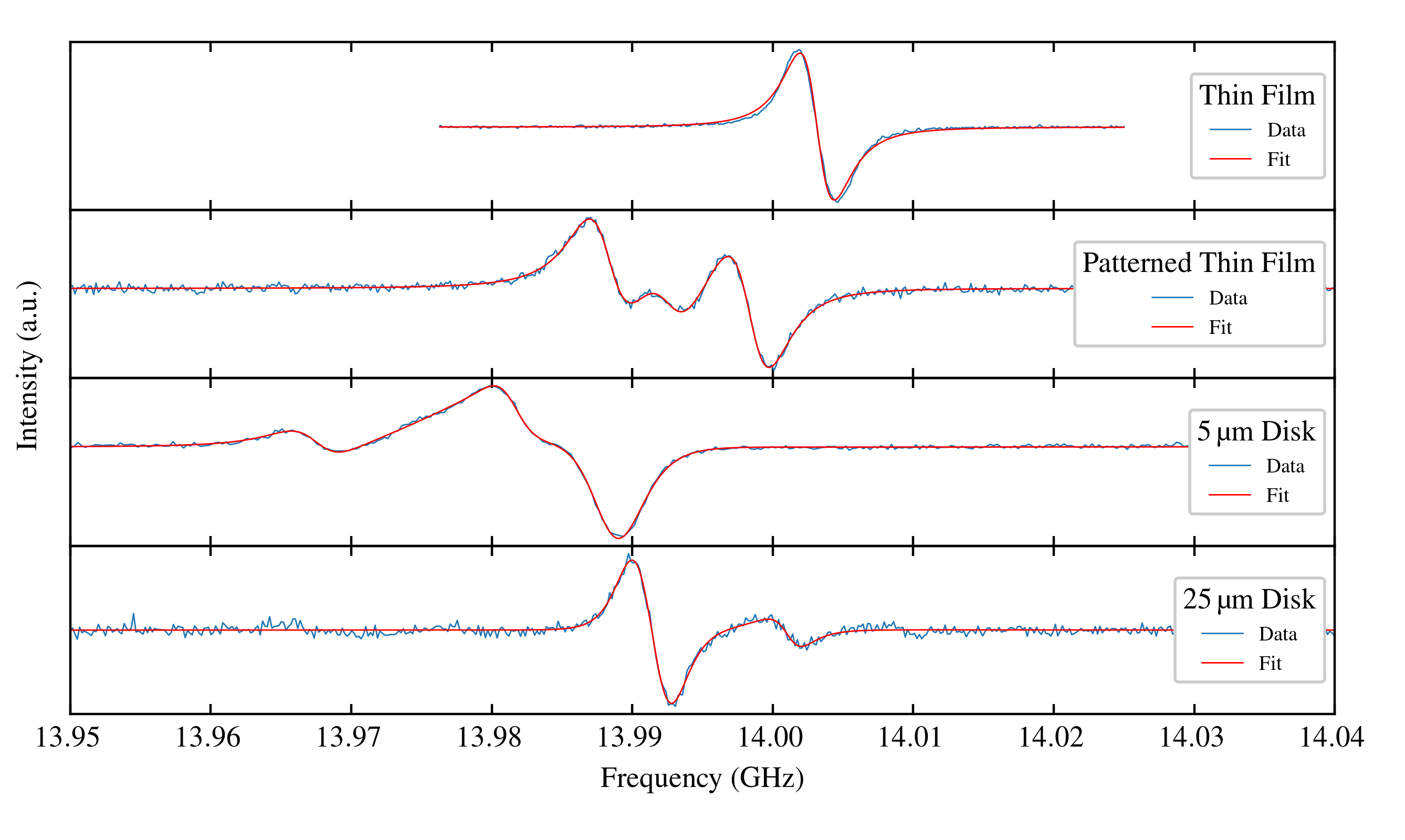}
  \caption{Representative  single  scans  of various  \VTCNE  features
      ranging from an unpatterned thin  film to a \SI{5}{\um} diameter
      disk  array.   All scans  are  done  in the  OOP  (\(\theta=0\))
      geometry. All  growths were 1-hour  long, resulting in a  300 nm
      thick film for the \SI{5}{\um} film and 400 nm thickness for the
      rest. The patterned thin film is a  2 mm by 2 mm patterned patch
      of  \VTCNE. The  linewidth  is  extracted from  the  fit to  the
      furthest  right lorentzian  except  for  the \SI{25}{\um}  disks
      where the linewidth is extracted from the largest peak.}
  \label{fig:BFMR}
\end{figure}

\newpage
Figure~\ref{fig:BFMR}  shows representative  linescans  from the  BFMR
measurement  setup. The  spectra  are  fit well  by  a combination  of
several lorentzian  derivatives.  Each lorentzian  derivative function
has    antisymmetric   (absorptive)    and   symmetric    (dispersive)
components\cite{Kalarickal_2006} represented by
\begin{subequations}
    \begin{equation}
        L'_{abs}(f) = \frac{a \Gamma^3 (f-f_0)}{\left(\Gamma^2 + 4 (f-f_0)^2\right)^2},
    \end{equation}
    \begin{equation}
        L'_{disp}(f) = \frac{-d \Gamma^2 \left(\Gamma^2-4 (f-f_0)^2\right)}{\left(\Gamma^2 + 4 (f-f_0)^2\right)^2},
    \end{equation}
\end{subequations}
so each derivative lorentzian in the fit is represented by
\begin{equation}
    L'_{total}(f) = \frac{a \Gamma^3 (f-f_0)}{\left(\Gamma^2 + 4 (f-f_0)^2\right)^2} -
    \frac{d \Gamma^2 \left(\Gamma^2-4 (f-f_0)^2\right)}{\left(\Gamma^2 + 4 (f-f_0)^2\right)^2}
\end{equation}
where  \(a\)  is  the  height  of the  derivative  of  the  absorptive
component, \(d\)  is the  height of the  derivative of  the dispersive
component,  \(\Gamma\)  is  the  full-width  at  half-maximum  of  the
lorentzian, \(f\) is the independent variable, and \(f_0\) is the peak
of the lorentzian.

\newpage
\section*{Analytic  Resonant field Calculations}

Here  we  find   an  analytical  expression  for   the  spin-wave  (or
magnetostatic mode) resonant fields for a normally magnetized cylinder
with  thickness \(d\)  and  radius \(R\).   We  first solve  Maxwell's
equations within the magnetostatic  regime.  Application of the proper
boundary conditions  at \(z=\nolinebreak\pm\frac{d}{2}\), the  top and
bottom surfaces  of our cylinder, yields  the following transcendental
equations\cite{Sparks_1970, denisflatte}
\begin{align}
\tan\left(k_{i}d\right) & =2\frac{k_{o}k_{i}}{k_{i}^{2}-k_{o}^{2}},\\
\frac{ik_{i}}{\sqrt{1+\kappa}} & =k_{o},\label{koki}
\end{align}
where  $k_{o}$ and  $k_{i}$  are the  in-plane  and out-of-plane  wave
vectors,                       respectively,                       and
$\kappa=\frac{\Omega_{H}}{\Omega_{H}^{2}-\Omega^{2}}$,            with
$\Omega_{H}=\frac{H+M_{s}\lambda_{ex}\left(k_{i}^{2}+k_{0}^{2}\right)}{M_{s}}$,
$\lambda_{ex}=\frac{2A_{ex}}{\mu_{0}M_{s}^{2}}$                    and
$\Omega=\frac{\omega}{\gamma M_{s}}$.
An  analytic expression  for the  resonant  spin-wave  fields is  then
obtained from  the Maxwell's  equation coupled to  the Laudau-Lifshitz
equation,  with the  additional assumption  that the  magnetization is
pinned at \(r=R\), which yields\cite{Sparks_1970, denisflatte}
\begin{align}
J_{m-1}\left(k_{o}R\right) & =0,\\
\rightarrow k_{o,m-1}^{n} & =\frac{\beta_{m-1}^{n}}{R},\label{bessel0}
\end{align}
where  \(\beta_{m-1}^{n}\) is  the  \symb{n^{th}}-zero  of the  Bessel
function   of   order   \(m-1\).    Now   using   Eq.~\ref{koki}   and
Eq.~\ref{bessel0} we obtain\cite{denisflatte} the following expression
for the spin-wave resonant fields
\begin{equation}
B_{nml}^{z}\approx\mu_{o}\frac{\omega}{\gamma}+\mu_{o}M_{s}-\frac{2A_{ex}}{M_{s}}\left[k_{i,nml}^{2}+\left(\frac{\beta_{m-1}^{n}}{R}\right)^{2}\right]-\frac{\mu_{o}M_{s}}{2\left(1+\frac{k_{i,nml}^{2}R^{2}}{\left(\beta_{m-1}^{n}\right)^{2}}\right)}-\frac{\mu_{o}\left(M_{s}\right)^{2}}{8\frac{\omega}{\gamma}\left(1+\frac{k_{i,nml}^{2}R^{2}}{\left(\beta_{m-1}^{n}\right)^{2}}\right)^{2}},\label{resonantfield}
\end{equation}
assuming
$\ensuremath{\frac{\omega}{\gamma}\gg
    M_{s}\left[2\left(1+\frac{k_{i,nml}^{2}R^{2}}{\left(\beta_{m-1}^{n}\right)^{2}}\right)\right]^{-1}}$. The
indices  \(n\), \(m\)  and \(l\)  represent the  radial, angular,  and
thickness  mode   numbers,  respectively.   The  resonant   fields  in
Eq.~\ref{resonantfield}  are  derived  using  the  SI  electromagnetic
equations and \(B_{nml}^{z}\)  has units of Tesla (T).   To obtain the
resonant  fields  \(H_{nml}^{z}\) in  Oersted  (Oe)  unit, the  values
obtained   from  Eq.~\ref{resonantfield}   should  be   multiplied  by
\(1\times 10^4\) Oe/T.

\begin{figure}[h]
    \begin{centering}
        \includegraphics[scale=0.375]{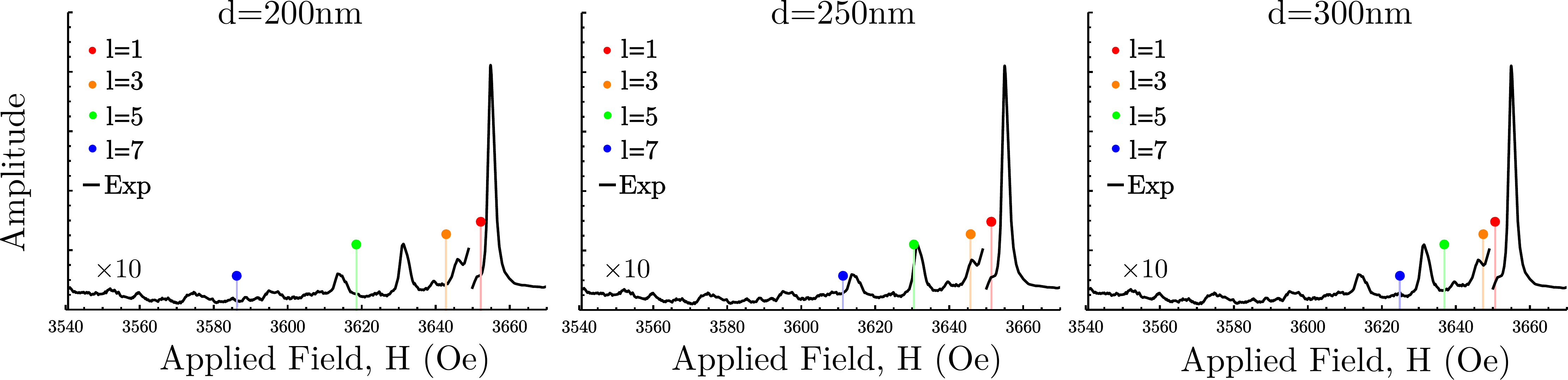}
        \par\end{centering}
      \caption{Plot  of  the  resonant   fields  for  the  first  five
        thickness modes  $l=1,3,5,7$ for the angular  and radial modes
        $m=1$ and $n=1$  a) $d=200$nm, b) $d=250$nm  and c) $d=300$nm.
        For all three  plots, $A=2.2\times \thinspace10^{-10}$~erg/cm,
        $M_{s}=76.57$~G, $N_{nml}\in[0.865-0.925]$, $\omega=9.83$ GHz,
        $\gamma=2.73\times 10^{6}$ MHz/Oe, and $R=2500$ nm.  The black
        line shows  the experimental spectrum obtained  for the \VTCNE
        cylinder   for   the   OOP  field   orientation   ($\theta=0$,
        $\varphi=0$).}
    \label{sfig:4}
\end{figure}

Eq.~\ref{resonantfield}  does  not  account  for  the  demagnetization
field. A good  match with the experimental data is  obtained using the
approached described  in Kakezei\cite{Kakazei_2004} that  considers an
effective demagnetization  per mode  \(N_{nml}\).  To account  for the
effect    of     the    demagnetization    field,     we    substitute
\(M_{s}           \rightarrow            M_{s}N_{nml}\)           with
\(N_{nml}<1\). Fig.~\ref{sfig:4} shows the resonant fields found using
Eq.~\ref{resonantfield}                                            for
\(N_{nml}\in[0.865-0.925]\)\cite{Kakazei_2004,Nedukh_2013,denisflatte}
for different \(d\). The best match  with the experimental data is for
\(d=250\)  nm.  This  is smaller  than  the nominal  thickness of  the
\VTCNE disks  used in the  experiment (300~nm), which we  attribute to
the fact that  the lens shape leads to a  smaller effective thickness.
Like the simulations, the analytical  calculations also predict a much
closer   spacing   for  the   radial   modes   as  compared   to   the
thickness-quantized modes (not shown).

\newpage
\nocite{*}
\bibliography{Suppbib}